\documentclass[prd,preprintnumbers,onecolumn,showkeys,floatfix,nofootinbib,notitlepage]{revtex4-1}
\usepackage{graphicx}
\usepackage{amssymb}
\usepackage{amsmath}
\usepackage{mathtools}
\usepackage{epsfig}
\usepackage{color}
\usepackage{enumerate}
\usepackage{slashed}
\usepackage{hyperref}

\def\slashchar#1{\setbox0=\hbox{$#1$}
   \dimen0=\wd0 \setbox1=\hbox{/} \dimen1=\wd1
   \ifdim\dimen0\big>\dimen1 \rlap{\hbox to \dimen0{\hfil/\hfil}} #1
   \else  \rlap{\hbox to \dimen1{\hfil$#1$\hfil}} / \fi}

\newcommand{\ud}{\mathrm{d}}
\newcommand{\be}{\begin{equation}}
\newcommand{\ee}{\end{equation}}
\newcommand{\bea}{\begin{eqnarray}}
\newcommand{\eea}{\end{eqnarray}}

\newcommand{\Appendix}[1]%
    {%
     \section{#1}%
      }

\begin{document}

\title{TMD-Factorization in Hadron-Hadron Collision}

\author{Gao-Liang Zhou
}
\affiliation{Key Laboratory of Frontiers in
Theoretical Physics, \\The Institute of Theoretical Physics, Chinese
Academy of Sciences, Beijing 100190, People's Republic of China}




\begin{abstract}
Proof of transverse-momentum-dependent(TMD) factorization for hadron-hadron collision is given in this paper. We focus on processes without detected soft final hadrons or detected final hadrons that are collinear to initial hadrons. This contradicts the widely accepted viewpoint that TMD-factorization does not hold in such processes even in the generalized sense. The key point is that singular points of the type $l^{+}=0$ can be absorbed into Wilson lines of soft gluons, where $l$ is collinear to the plus direction. Thus one should subtract such singular points from the collinear region of $l$. After such subtraction, one can make use of Ward identity to absorb effects of scalar-polarized collinear gluons into Wilson lines.
\end{abstract}

\maketitle

\section{Introduction}
\label{introduction}

Transverse-momentum-dependent(TMD) factorization in hadron-hadron collisions with large transverse momentum back-to-back hadrons produced in the final states is a much non-trivial issue. (\cite{BMP:2004,BBMP:2005,BMP:2006,BM:2008,CQ:2007,VY:2007,C:2007,BBAMM:2007,BMVY:2007,BMP:2008,RM:2010,Roggers:2013})
This is because that the Ward-identity cancelation is prevented by the singular points of the type $l^{+}=0$, where $l^{\mu}$ is collinear to the plus direction.
It was first noted in \cite{BMP:2004,BMP:2006} that singular points of such type can cause the
process-dependence(or diagram-dependence) in Wilson-line structures of TMD parton distribution functions(PDFs) and TMD fragmentation functions(FFs). According to calculations in a model theory, authors in \cite{CQ:2007,C:2007} found that the process-dependent Wilson-lines can cause the break down of normal proofs of standard factorization in the inclusive production of two high-transverse-momentum hadrons in hadron-hadron collisions. It was later found in \cite{RM:2010} that TMD-factorization in  production of back-to-back hadrons in hadron-hadron collisions fails even one allows Wilson-lines in TMD-functions to be process dependent. In addition to the anomalous color caused by the process dependent Wilson lines, the nontrivial dependence of cross sections on transverse momenta of gluons exchanged between spectators and active partons is also identified in \cite{Roggers:2013} as constituting a breakdown of TMD-factorization.

Unlike the case in QCD process, TMD-factorization theorems have been derived rigorously in perturbative QCD(pQCD) for a number of electromagnetic processes including
Drell-Yan (DY) process, semi-inclusive deep inelastic scattering (SIDIS) process  and the production of back-to-back hadrons in $e^{+}e^{-}$ annihilation process.(\cite{CS:1981B,CS:1982,B:1985,CSS:1985,CSS:1988,JMY:2004,JMY:2004,JMY:2005,JMY:2005,pQCD:2011})
Even for such processes, one should first deform the integral paths of gluon momenta so that the Grammer-Yennie approximation and the Ward-identity cancelation works(\cite{B:1985,CSS:1985,CSS:1988}). The deformations are different for various processes. For SIDIS and $e^{+}e^{-}$ annihilation the deformations should be consistent with exchanges of gluons between spectators and final particles. While for DY the deformations should be consistent with those between spectators and initial particles.(\cite{CM:2004}) This makes the Wilson lines in SIDIS and $e^{+}e^{-}$ annihilation to be future-pointing and those in DY to be past pointing. Such differences can cause a sign flip for the Sivers function in DY as compared to SIDIS.(\cite{Collions:2002}) Thus TMD-functions can be process dependent even for such processes, although the process dependence can be absorbed into future pointing or past pointing Wilson lines.  Relations between these functions in SIDIS, $e^{+}e^{-}$ and DY are further studied in \cite{BMP:2003}. Other issues about the Wilson lines in these processes include, for example, transverse Wilson lines at infinity(\cite{BJY:2003}), rapidity divergences
(see, for example, Refs.\cite{CH:2000,Collins:2003,Hautmann:2007,Collins:2008}) and cusp structures of Wilson lines(see, for example, Refs.\cite{CS:2008,Cherednikov:2012}).

The deformation of integral path is closely related to the issue of Glauber gluons(\cite{BBL:1981}), which are space-like soft gluons with the momenta $q^{\mu}$ locate in the region:
$q^{0}\ll |\vec{q}|\sim \Lambda_{QCD}$. In the coupling between a soft gluon with momentum $q$(denoted as $A_{s}^{\mu}(q)$) and a particle collinear-to-plus with momentum $p$ and mass $m$, one make the Grammer-Yennie approximation(\cite{GY:1973}):
\begin{equation}
A_{s}^{\mu}(q)\to A_{s}^{-}(q)
,\quad
(p+q)^{2}\to p^{2}-m^{2}+2p^{+}q^{-}
\end{equation}
so that the soft gluon $q$ behave as a scalar polarized gluon in such coupling. One can then make use of Ward identity to absorb this gluon into a Wilson line along the plus direction. Gluons in Glauber region, however, break such approximation. For example, one may consider the case that $q^{\mu}$ locate in the region: $(q^{+},q^{-},|\vec{q}_{\perp}|)\sim \Lambda_{QCD}(\Lambda_{QCD}/p^{+},\Lambda_{QCD}/p^{+},1)$. In this case, we have:
\begin{equation}
p^{2}-m^{2}+2p^{+}q^{-}\sim q^{2}-2\vec{p}_{\perp}\cdot \vec{q}_{\perp}\sim \Lambda_{QCD}^{2}
\end{equation}
Thus the Grammer-Yennie approximation does not work in this case. To make the Grammer-Yennie approximation work, one should first deform the integral paths out of the Glauber region(see, for example, Refs.\cite{S:1978,CS:1981S,B:1985,CSS:1985,CSS:1988,CSS:1989}) Deformations of the integral paths depend on wether interactions between soft gluons and collinear particles occur before or after the hard collision. There are not universal deformation of integral paths to avoid Glauber in hadron-hadron collisions as both initial and final states interactions exist in such process. It is well known that standard factorization does not hold for hadron-hadron collision with detected hadrons in the target momenta region.(\cite{CFS:1993,CDF:1997,ACTW:1997,G:1997,H1:2007}) For inclusive hadron-hadron process, like the Drell-Yan process, final poles in Glauber region cancel out. Thus one can deform the integral paths to be consistent with those between spectators and initial particles so that factorization holds for such processes.(\cite{B:1985,CSS:1985,CSS:1988})

The issue of Glauber gluons is also related to collinear factorizations, in which transverse momenta of partons are integrated over in the definitions of PDFs and FFs. In this case, factorization is saved by cancelations of poles of final states interactions once integral over transverse momenta is performed.(\cite{NQS:2005,AS:2009})  For hadron-hadron collisions with large transverse momentum back-to-back hadrons, however, the cross section is not sufficiently inclusive in transverse momenta. Thus the cancellation does not simply work in such processes.

In this paper, we present an approach to show that the similar cancellation does occur for the hadron-hadron collisions with large transverse momentum back-to-back hadrons produced in the final state. The physical picture of such proof is clear. Couplings between soft gluons and final jets produced in the hard-subprocess can be described by the Grammer-Yennie approximation as such couplings always occurs after the hard scattering. After this approximation, soft gluons couple to eikonal lines along the directions of final jets instead of couple to final jets direct. We can thus factorize the detected jets from the scattering matrix-elements. This is represented by the formula:
\begin{equation}
M=(\prod_{i=1}^{2}M_{det}^{i})M_{inc}
\end{equation}
where $M$ represents a diagram that contribute to the process, $M_{det}^{i}$ represent the collinear subgraph of the $i$-th jets. $M_{inc}$ includes the hard subgraph, soft subgraph, collinear subgraph of other jets and Wilson lines along the directions of the detected jets. Coherence between $M_{def}^{i}$ and $M_{inc}$ can only be caused by coherence between their spin structures,  color structures and total momenta. Hadrons are color singlets, thus the color charge of $M_{det}^{i}$ should be the same as that of its conjugation. One should perform the summation over all possible final particles with definite total momentum, color charge and spin in $M_{inc}M_{inc}^{*}$. The action of QCD is colorless and Poincar$\acute{e}$ invariant. Thus such states form the invariant sub-space of QCD. After such summation, couplings between soft gluons and collinear particles after the hard collision cancel out in $M_{inc}M_{inc}^{*}$. We can then deform
the integral paths to avoid Glauber region in $M_{inc}M_{inc}^{*}$. The Grammer-Yennie works in $M_{inc}M_{inc}^{*}$ after this deformation.

We then absorb soft gluons into Wilson lines along the directions of collinear particles. We subtract contributions of these Wilson lines from couplings between gluons and spectators. After such subtraction, We can take the collinear approximation in the coupling between active partons and such gluons. Singular points of the type $l^{+}=0$ do not give leading order contribution after this subtraction, where $l^{\mu}$ represents the momenta of gluons that are is collinear-to-plus. We can then make use of Ward identity to absorb collinear gluons into past pointing or future pointing Wilson lines.

The paper is organized as follows. We first consider a simple example in Sec.{\ref{cancellation}}, in which one gluon is exchanged between spectators and active partons or spectators. According to explicit calculations,  we show the cancelation of pinch singular points in Glauber region in this example. We then prove such cancelation in the frame of effective theory in Sec.\ref{deformation}. Wilson lines of soft gluons and collinear gluons are brought in in Sec.\ref{Wilson lines}. The effective vertex of hard subprocess is also construct in this section. In Sec.\ref{tr dependence}, we consider the same process as in \cite{Roggers:2013} to show that the nontrivial transverse momenta dependence of cross section can be described by the effective hard vertex brought in Sec.\ref{Wilson lines}. We finish the proof factorization theorem in Sec.\ref{factorize}. Some discussions are given in Sec.\ref{conclusion}.
\\

\section{Cancellation of Pinch Singularities in Glauber Region at Order $\alpha_{s}$}
\label{cancellation}

In this section, we show the cancellation of leading pinch singular surfaces(LPSS) in Glauber region in a simile example. There is one gluon exchanged between spectators and active partons or spectators in the graphs considered in this section. We work in Feynman gauge in this section.

The process we considered in this paper can be written as:
\begin{equation}
A(p_{1})+B(p_{2})\to H_{3}(k_{3})+H_{4}(k_{4})+X
\end{equation}
where $A$ and $B$ represent (polarized or unpolarized)initial hadrons with momenta $p_{1}$ and $p_{2}$, $H_{3}$ and $H_{4}$ represent detected back-to-back (polarized or unpolarized)hadrons with momenta $k_{3}$ and $k_{4}$, $X$ represents any other states.

We work in the center of mass frame of initial hadrons. $p_{1}$ and $p_{2}$ are taken as nearly collinear to plus and minus directions respectively. That is:
\begin{equation}
p_{1}=(p_{1}^{+},M_{1}^{2}/2p_{1}^{+},\vec{0})\simeq (p_{1}^{+},0,\vec{0})
\end{equation}
\begin{equation}
p_{2}=(M_{2}^{2}/2p_{2}^{-},p_{2}^{-},,\vec{0})\simeq (0,p_{2}^{-},\vec{0})
\end{equation}
where $M_{1}$ and $M_{2}$ denote the masses of initial hadrons, which are of order $\Lambda_{QCD}$. We bring in the notation:
\begin{equation}
\label{n+}
n_{1}^{\mu}=\frac{1}{\sqrt{2}}(1,0,0,1),\quad
n_{2}^{\mu}=\frac{1}{\sqrt{2}}(1,0,0,-1)
\end{equation}
\begin{equation}
\bar{n_{i}}^{\mu}=\sqrt{2}(1,0,0,0)-n_{i}^{\mu}
\end{equation}
and have:
\begin{equation}
p_{i}^{\mu}=\bar{n_{i}}\cdot p_{i}n_{i}^{\mu}(1+O(M_{i}/Q))
\end{equation}
for $i=1,2$.
We assume that $k_{3}$ and $k_{4}$ nearly collinear to the direction $n_{3}^{\mu}$ and $n_{4}^{\mu}$ respectively, where:
\begin{equation}
n_{3}^{\mu}=\frac{1}{\sqrt{2}}(1,\sin(\theta_{3})\cos(\phi_{3}),
\sin(\theta_{3})\sin(\phi_{3}),\cos(\theta_{3}))
\end{equation}
\begin{equation}
n_{4}^{\mu}=\frac{1}{\sqrt{2}}(1,\sin(\theta_{4})\cos(\phi_{4}),
\sin(\theta_{4})\sin(\phi_{4}),\cos(\theta_{4}))
\end{equation}
with $1-\cos(\theta_{3})\sim 1+\cos(\theta_{3})\sim 1$,
$1-\cos(\theta_{4})\sim 1+\cos(\theta_{4})\sim 1$, $\pi-(\theta_{3}+\theta_{4})\sim 0$ and $\pi-|\phi_{3}-\phi_{4}|\sim 0$.
We then have:
\begin{equation}
k_{i}^{\mu}=\bar{n_{i}}\cdot k_{i}n_{i}^{\mu}(1+O(M_{i}/Q))
\end{equation}
where
\begin{equation}
\bar{n_{i}}^{\mu}=\sqrt{2}(1,0,0,0)-n_{i}^{\mu}
\end{equation}
for $i=3,4$. We also bring in the notation:
\begin{equation}
q^{\mu}=k_{3}^{\mu}+k_{4}^{\mu},\quad Q=\sqrt{q^{2}}
\end{equation}
Then the hard energy scale of the process is set by $Q$.

\begin{figure*}
\begin{tabular}{c@{\hspace*{10mm}}c}
\includegraphics[scale=0.3]{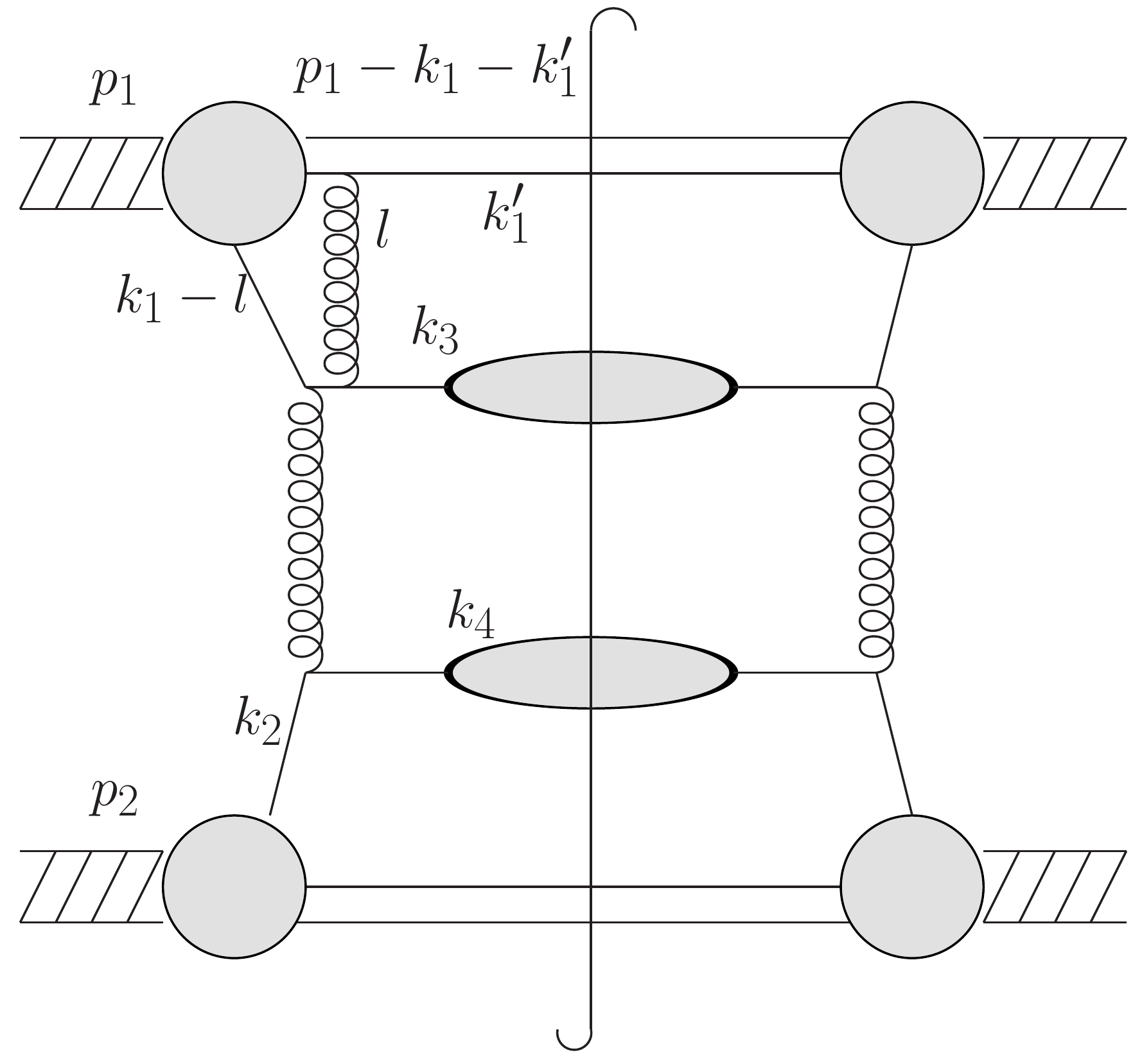}
&
\includegraphics[scale=0.3]{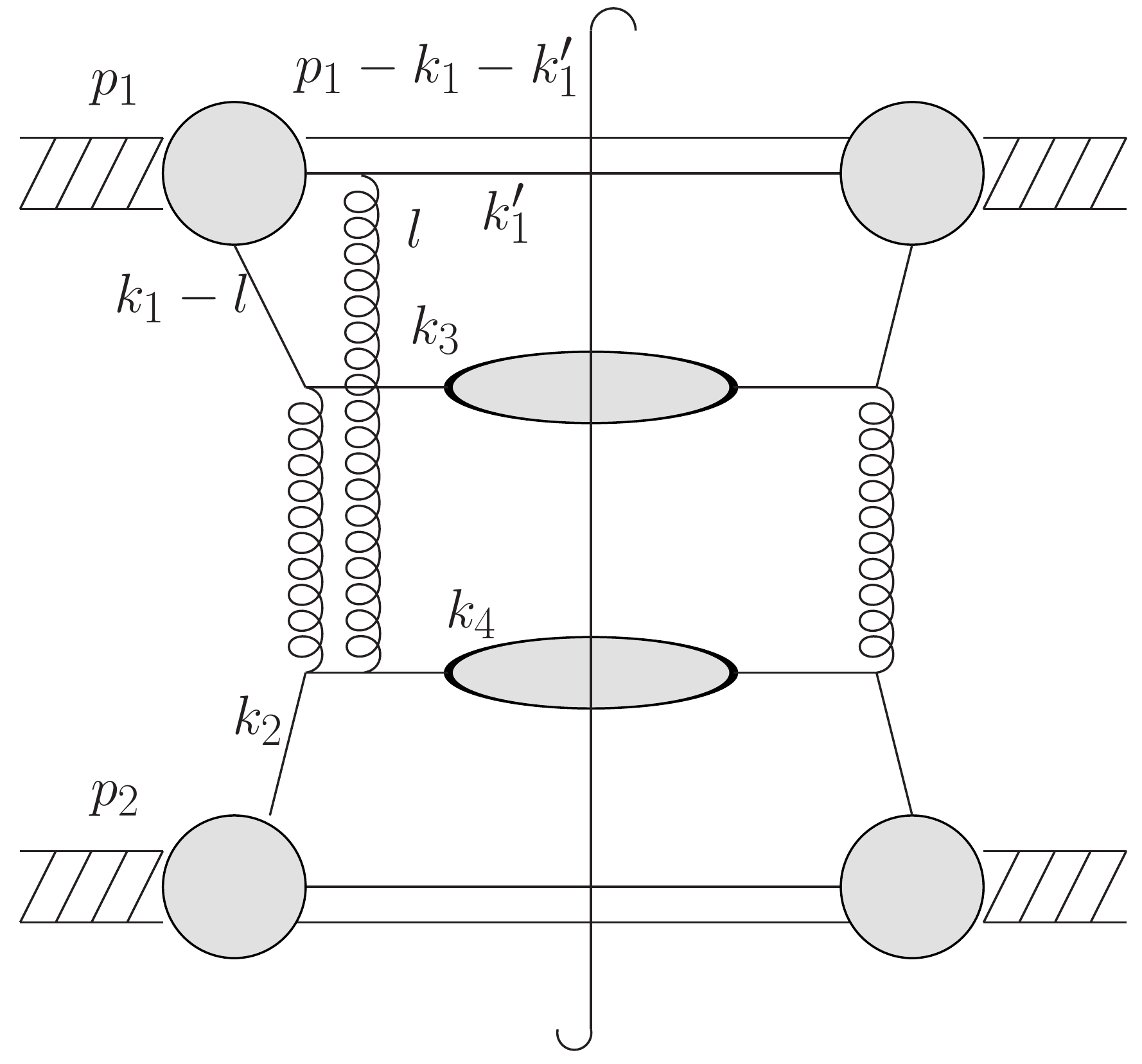}
\\
(a)&(b)
\\
\includegraphics[scale=0.3]{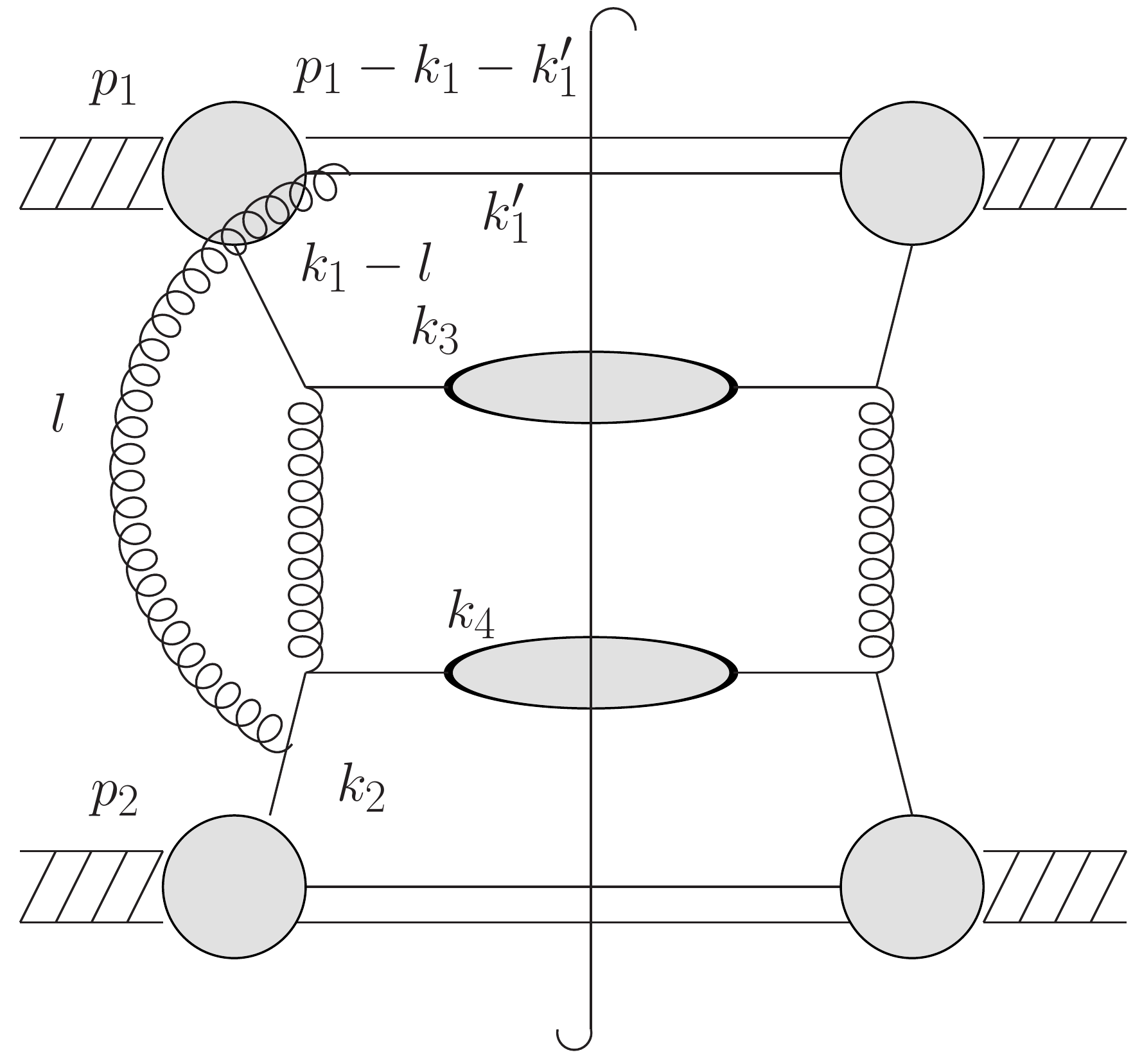}
&
\includegraphics[scale=0.3]{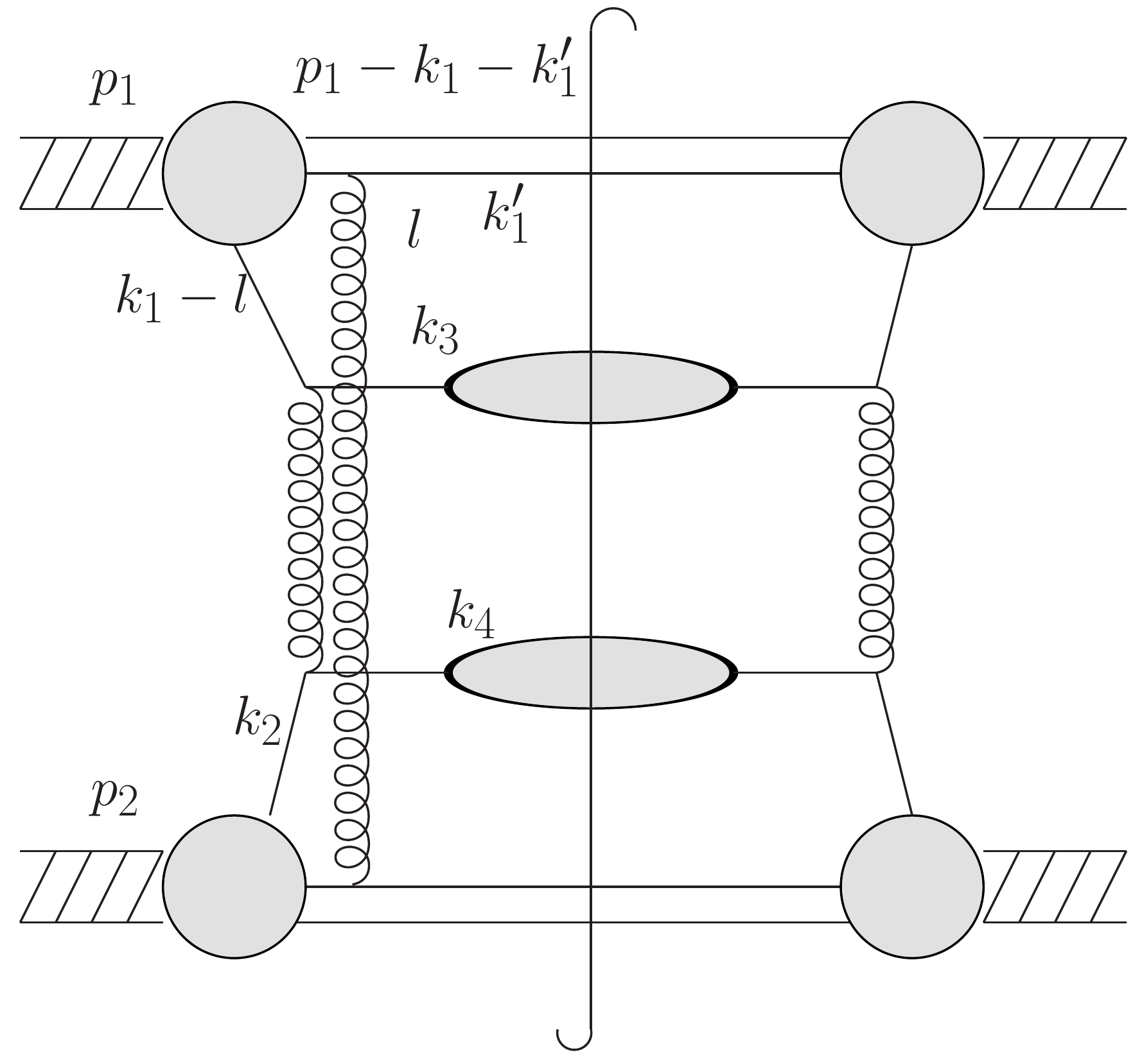}
\\
(c)&(d)
\end{tabular}
\caption{Diagrans with one gluon exchanged between remnants of $p_{1}$ and other particles}
\label{subtraction}
\end{figure*}

We consider the diagrams with one gluon exchanged between remnants of $p_{1}$ and other particles. These are diagrams shown in Fig.(\ref{subtraction}) and their conjugations.

We start from the first diagram in Fig.\ref{subtraction}.
The part of this diagram that depends on the momentum $l$ can be written as:
\begin{eqnarray}
\label{l dependent}
M_{1a;\rho b}^{i_{1}i_{3}}(l)&\equiv&(-i)\left(\gamma^{\mu}t^{a}\frac{\not\! k_{1}^{\prime}+\not\! l}{(k_{1}^{\prime}+l)^{2}+i\epsilon}\right)_{i_{1}j_{1}}
\nonumber\\
&&
\Gamma_{H}^{j_{1}j_{3}}(p,k_{1}-l,k_{1}^{\prime}+l)
\frac{g^{\mu\nu}}{l^{2}+i\epsilon}
\nonumber\\
&&
\left(t^{a}\gamma^{\nu}
\frac{\not\! k_{3}-\not\! l}{(k_{3}-l)^{2}+i\epsilon}
\gamma^{\rho}t^{b}
\frac{\not\! k_{1}-\not\! l}{(k_{1}-l)^{2}+i\epsilon}\right)_{i_{3}j_{3}}
\end{eqnarray}
where
$\Gamma_{H}$ denotes the effective vertex of couplings between partons and hadrons, $i_{1}$ and $i_{3}$ denote the Dirac indices of the spinor $\bar{u}(k_{1})$ and $\bar{u}(k_{3})$, $j_{1}$ and $j_{3}$ denote the Dirac indices of the vertex $\Gamma_{H}$(there may be more Dirac indices in $\Gamma_{H}$, we do not display them explicitly), $\rho$ and $b$ represents the Lorentz and color indices of the hard gluon, the factor $-i$ is that relative to the diagram without gluons.  $l$ is  collinear-to-plus or soft(including Glauber) at leading order according to infrared power counting.(\cite{S:1978}).

We first consider the case that $l$ locate in the soft region. In this region,  we can write $M_{1a;\rho b}(l)$ at leading order as:
\begin{eqnarray}
&&(-i)t^{a}\frac{2k_{1}^{\prime+}}
{(k_{1}^{\prime}+l)^{2}+i\epsilon}
\Gamma_{H}^{i_{1}j_{3}}(p,k_{1}-l,k_{1}^{\prime}+l)
\frac{1-\cos(\theta_{3})}{2(l^{2}+i\epsilon)}
\nonumber\\
&&
\left(t^{a}
\frac{2\bar{n_{3}}\cdot k_{3}}{(k_{3}-l)^{2}+i\epsilon}
\gamma^{\rho}t^{b}
\frac{k_{1}^{+}\gamma^{-}}{(k_{1}-l)^{2}+i\epsilon}\right)_{i_{3}j_{3}}
\end{eqnarray}
Poles of the internal lines collinear-to-plus, the internal line collinear to $n^{\mu}$ and soft internal line  read $l^{-}\gtrsim M_{1}^{2}/p_{1}^{+}$, $n_{3}\cdot l\gtrsim M_{3}^{2}/\bar{n_{3}}\cdot k_{3}$ and $l^{0}\sim E_{\vec{l}}$ respectively. We have assumed that transverse components of $k_{3}$ are large, thus $n_{3}\cdot l$ do not pinch in Glauber region. We can deform the integral path of $n_{3}\cdot l$ to lower half plane to avoid the Glauber region and  take Grammer-Yennie approximation in the internal line $k_{3}-l$:
\begin{equation}
\frac{1}{(k_{3}-l)^{2}+i\epsilon}\simeq \frac{1}
{-2\bar{n_{3}}\cdot k_{3}n_{3}\cdot l+i\epsilon}
\end{equation}
After this approximation, we deform the integral path of $n_{3}\cdot l$ back to real axis and define the quantity:
\begin{eqnarray}
\widetilde{M}_{1a;\rho b}^{i_{1}i_{3}}
&\equiv&(-i)t^{a}\frac{2k_{1}^{\prime+}}
{(k_{1}^{\prime}+l)^{2}+i\epsilon}
\Gamma_{H}^{i_{1}j_{3}}(p,k_{1}-l,k_{1}^{\prime}+l)
\frac{1-\cos(\theta_{3})}{2(l^{2}+i\epsilon)}
\nonumber\\
&&
\left(t^{a}
\frac{1}{-n_{3}\cdot l+i\epsilon}
\gamma^{\rho}t^{b}
\frac{k_{1}^{+}\gamma^{-}}{(k_{1}-l)^{2}+i\epsilon}\right)_{i_{3}j_{3}}
\end{eqnarray}
We then make the decomposition:
\begin{equation}
M_{1a;\rho b}(l)=\widetilde{M}_{1a;\rho b}(l)+M_{1a;\rho b}(l)-\widetilde{M}_{1a;\rho b}(l)=\widetilde{M}_{1a;\rho b}(l)+\widehat{M}_{1a;\rho b}(l)
\end{equation}
We see that the $\widehat{M}_{1a;\rho b}(l)$ part is power suppressed in the soft region.

We now consider contributions of the momentum region:
\begin{equation}
|\vec{l}_{\perp}|\sim M_{1},\quad
|l^{-}|\ll |\vec{l}_{\perp}|,\quad
|l^{+}|\lesssim |\vec{l}_{\perp}|
\end{equation}
The $\widehat{M}_{1a;\rho b}(l)$ part part is power suppressed in this region and can be neglected. We consider the summation of the first diagram in Fig.\ref{subtraction} and its conjugation. The parts of these two diagrams that depend on $\vec{l}_{\perp}$ can be written as:
\begin{eqnarray}
\sigma_{1a;\rho b;\rho^{\prime}b^{\prime}}^{i_{1}i_{3}i_{1}^{\prime}i_{3}^{\prime}}(\vec{l}_{\perp})&\equiv&
\int\frac{\ud l^{+}}{2\pi}\frac{\ud l^{-}}{2\pi} (-i)\Gamma_{H}^{*i_{1}^{\prime}j_{3}^{\prime}}(p_{1},k_{1},k_{1}^{\prime})
\left(t^{b^{\prime}}
\frac{k_{1}^{+}\gamma^{-}}{k_{1}^{2}-i\epsilon}
\gamma^{\rho^{\prime}}\right)_{j_{3}^{\prime}i_{3}^{\prime}}
\nonumber\\
&&
t^{a}\frac{2k_{1}^{\prime+}}
{(k_{1}^{\prime}+l)^{2}+i\epsilon}
\Gamma_{H}^{i_{1}j_{3}}(p,k_{1}-l,k_{1}^{\prime}+l)
\frac{1-\cos(\theta_{3})}{2(l^{2}+i\epsilon)}
\nonumber\\
&&
\left(t^{a}
\frac{1}{-n_{3}\cdot l+i\epsilon}
\gamma^{\rho}t^{b}
\frac{k_{1}^{+}\gamma^{-}}{(k_{1}-l)^{2}-i\epsilon}\right)_{i_{3}j_{3}}
\nonumber\\
&&+\int\frac{\ud l^{+}}{2\pi}\frac{\ud l^{-}}{2\pi} (i)
t^{a}\frac{2k_{1}^{\prime+}}
{(k_{1}^{\prime}+l)^{2}+i\epsilon}
\nonumber\\
&&
\Gamma_{H}^{*i_{1}^{\prime}j_{3}^{\prime}}(p,k_{1}-l,k_{1}^{\prime}+l)
\frac{1-\cos(\theta_{3})}{2(l^{2}-i\epsilon)}
\nonumber\\
&&
\left(t^{a}
\frac{1}{-n_{3}\cdot l-i\epsilon}
t^{b^{\prime}}
\frac{k_{1}^{+}\gamma^{-}}{(k_{1}-l)^{2}-i\epsilon}\gamma^{\rho^{\prime}}
\right)_{j_{3}^{\prime}i_{3}^{\prime}}
\nonumber\\
&&
\Gamma_{H}^{i_{1}j_{3}}(p_{1},k_{1},k_{1}^{\prime})
\left(t^{b}\gamma^{\rho}\frac{k_{1}^{+}\gamma^{-}}
{k_{1}^{2}+i\epsilon}\right)_{i_{3}^{\prime}j_{3}^{\prime}}
\end{eqnarray}
We integrate out $l^{-}$ by take the residue of the singular point:
\begin{equation}
\label{l-}
l^{-}=\frac{(\vec{k_{1}^{\prime}}_{\perp}+\vec{l}_{\perp})^{2}}
{2(k_{1}^{\prime +}+l_{1}^{+})}-k_{1}^{\prime-}-i\epsilon
\simeq \frac{(\vec{k_{1}^{\prime}}_{\perp}+\vec{l}_{\perp})^{2}}
{2k_{1}^{\prime +}}-k_{1}^{\prime-}-i\epsilon
\end{equation}
and its conjugation. Initial hadrons are stable states, thus $k_{1}^{\prime}+l$ and $k_{1}-l$ can not be both on-shell. We then integrate out $l^{+}$ by take the the residue of the singular point:
\begin{equation}
\label{l+}
l^{+}\simeq \frac{\vec{n_{3}}\cdot\vec{l}_{\perp}}{n_{3}^{-}}+i\epsilon
\end{equation}
and its conjugation. We do not take the residue of the singular point $l^{2}=0$ as such singular point do not locate in the region considered here. We then have:
\begin{eqnarray}
\sigma_{1a;\rho b}^{i_{1}i_{3}i_{1}^{\prime}i_{3}^{\prime}}(\vec{l}_{\perp})&\simeq&
(-i)\Gamma_{H}^{*i_{1}^{\prime}j_{3}^{\prime}}(p_{1},k_{1},k_{1}^{\prime})
\left(\frac{k_{1}^{+}\gamma^{-}}{k_{1}^{2}}\gamma^{\rho\prime}
t^{b^{\prime}}\right)_{j_{3}^{\prime}i_{3}^{\prime}}
\nonumber\\
&&
t^{a}
\Gamma_{H}^{i_{1}j_{3}}(p,k_{1}-l,k_{1}^{\prime}+l)
\frac{1}{|\vec{l}_{\perp}|^{2}}
\left(t^{a}
\gamma^{\rho}t^{b}
\frac{k_{1}^{+}\gamma^{-}}{(k_{1}-l)^{2}}\right)_{i_{3}j_{3}}
\nonumber\\
&&+(i)
t^{a}
\Gamma_{H}^{*i_{1}^{\prime}j_{3}^{\prime}}(p,k_{1}-l,k_{1}^{\prime}+l)
\frac{1}{|\vec{l}_{\perp}|^{2}}
\left(t^{a}t^{b\prime}
\frac{k_{1}^{+}\gamma^{-}}{(k_{1}-l)^{2}}
\gamma^{\rho^{\prime}}\right)_{j_{3}^{\prime}i_{3}^{\prime}}
\nonumber\\
&&
\Gamma_{H}^{i_{1}j_{3}}(p_{1},k_{1},k_{1}^{\prime})
\left(\gamma^{\rho}\frac{k_{1}^{+}\gamma^{-}}{k_{1}^{2}}\right)_{i_{3}j_{3}}
\nonumber\\
&=&
(-i)\frac{C_{f}t^{b}t^{b^{\prime}}}{|\vec{l}_{\perp}|^{2}}
\frac{k_{1}^{+}}{k_{1}^{2}}
\frac{k_{1}^{+}}{(k_{1}-l)^{2}}
(\gamma^{-}\gamma^{\rho^{\prime}})_{j_{3}^{\prime}i_{3}^{\prime}}
(\gamma^{\rho}\gamma^{-})_{i_{3}j_{3}}
\nonumber\\
&&
\left\{
\Gamma_{H}^{*i_{1}^{\prime}j_{3}^{\prime}}(p_{1},k_{1},k_{1}^{\prime})
\Gamma_{H}^{i_{1}j_{3}}(p,k_{1}-l,k_{1}^{\prime}+l)
\right.
\nonumber\\
&&
\left.
-
\Gamma_{H}^{*i_{1}^{\prime}j_{3}^{\prime}}(p,k_{1}-l,k_{1}^{\prime}+l)
\Gamma_{H}^{i_{1}j_{3}}(p_{1},k_{1},k_{1}^{\prime})
\right \}
\end{eqnarray}
where $l^{+}$ and $l^{-}$ are determined by (\ref{l+}) and (\ref{l-}), $C_{f}=\frac{N_{c}^{2}-1}{2N_{c}}$. We have made use of the colorlessness of $\Gamma_{H}$.  $k_{1}^{\prime}+l$ and $k_{1}-l$ can not be both on-shell as the initial hadron $p_{1}$ is stable particle. We thus drop the $i\epsilon$ terms in $(k_{1}-l)^{2}\pm i\epsilon$. We drop such terms in $k_{1}^{2}\pm i\epsilon$ according to the same reason. We also drop such terms in $l^{2}\pm i\epsilon$ as $l$ can not be on-shell in the region we considered here.

$\Gamma_{H}$ is Hermitian in both momenta space and spinor space, we thus have:
\begin{equation}
\Gamma_{H}^{*i_{1}^{\prime}j_{3}^{\prime}}(p,k_{1}-l,k_{1}^{\prime}+l)
\Gamma_{H}^{i_{1}j_{3}}(p_{1},k_{1},k_{1}^{\prime})
=\Gamma_{H}^{*i_{1}^{\prime}j_{3}^{\prime}}(p_{1},k_{1},k_{1}^{\prime})
\Gamma_{H}^{i_{1}j_{3}}(p,k_{1}-l,k_{1}^{\prime}+l)
\end{equation}
We then have:
\begin{eqnarray}
\sigma_{1a;\rho b}^{i_{1}i_{3}i_{1}^{\prime}i_{3}^{\prime}}(\vec{l}_{\perp})&\simeq&
0
\end{eqnarray}
in the region considered here.
Thus we can deform the integral path of $l^{-}$ to the lower half plane with radius of order $\min\{|\vec{l}_{\perp}|,\frac{|\vec{l}_{\perp}|^{2}}{|l^{+}|}\}$ to avoid the region:
\begin{equation}
|\vec{l}_{\perp}|\sim M_{1},\quad
|l^{-}|\ll |\vec{l}_{\perp}|,\quad
|l^{+}|\lesssim |\vec{l}_{\perp}|
\end{equation}
in the summation of first diagram of Fig.\ref{subtraction} and its conjugation.

For the second and third diagrams of Fig.\ref{subtraction}, we have the same conclusion. The fourth diagram of Fig.\ref{subtraction} is the same as that in Drell-Yan process. Thus we can also deform the integral path of $l^{-}$ to the lower half plane with radius of order $\min\{|\vec{l}_{\perp}|,\frac{|\vec{l}_{\perp}|^{2}}{|l^{+}|}\}$ so that the Grammer-Yennie approximation works in the coupling between $l$ and remnants of the initial hadron $p_{1}$.

We now consider the integration of $M_{1a;\rho b}^{i_{3}j_{3}}(l)$ over $l$. We deform the integral path of $l^{-}$ to the lower half plane with radius of order $\min\{|\vec{l}_{\perp}|,|\vec{l}_{\perp}|^{2}/|l^{+}|\}$. Residues of the singular points confronted in such deformation do not contribute to the whole process at leading order in $\Lambda_{QCD}/Q$ and can be dropped. We get the integral:
\begin{eqnarray}
\label{Meff}
&&\int \frac{\ud l^{+}}{2\pi}\int\frac{\ud^{2}\vec{l}_{\perp}}{(2\pi)^{2}}
\int_{C(l^{+},l_{\perp})}\frac{\ud l^{-}}{2\pi}M_{1a;\rho b}^{i_{1}i_{3}}(l)
\end{eqnarray}
where $C(l^{+},l_{\perp})$ represents the integral path of $l^{-}$ depending on $l^{+}$ and $\vec{l}_{\perp}$. In the soft region of $l$, we can make the approximation:
\begin{eqnarray}
M_{1a;\rho b}^{i_{1}i_{3}}(l)
&\simeq&
S_{1a;\rho b}^{i_{1}i_{3}}(l)
\nonumber\\
&\equiv& C_{f}t^{b}
\left(\gamma^{+}\frac{\gamma^{-}}{2l^{-}+i\epsilon}\right)_{i_{1}j_{1}}
\Gamma_{H}^{j_{1}j_{3}}\frac{1-\cos(\theta_{3})}{2(l^{2}+i\epsilon)}
\nonumber\\
&&
\left(\frac{k_{1}^{+}\gamma^{-}}{k_{1}^{2}-2k_{1}^{+}l^{-}+i\epsilon}\gamma^{\rho}
\frac{\not\!n}{-2n\cdot l+i\epsilon}
t^{a}\bar{\not\! n}\right)_{i_{3}j_{3}}
\end{eqnarray}
We then make the decomposition:
\begin{equation}
\label{decomposition}
M_{1a;\rho b}^{i_{1}i_{3}}(l)=S_{1a;\rho b}^{i_{1}i_{3}}(l)
+M_{1a;\rho b}^{i_{1}i_{3}}(l)-S_{1a;\rho b}^{i_{1}i_{3}}(l)
\equiv S_{1a;\rho b}^{i_{1}i_{3}}(l)
+C_{1a;\rho b}^{i_{1}i_{3}}(l)
\end{equation}

\begin{figure*}
\centering
\includegraphics[scale=0.3]{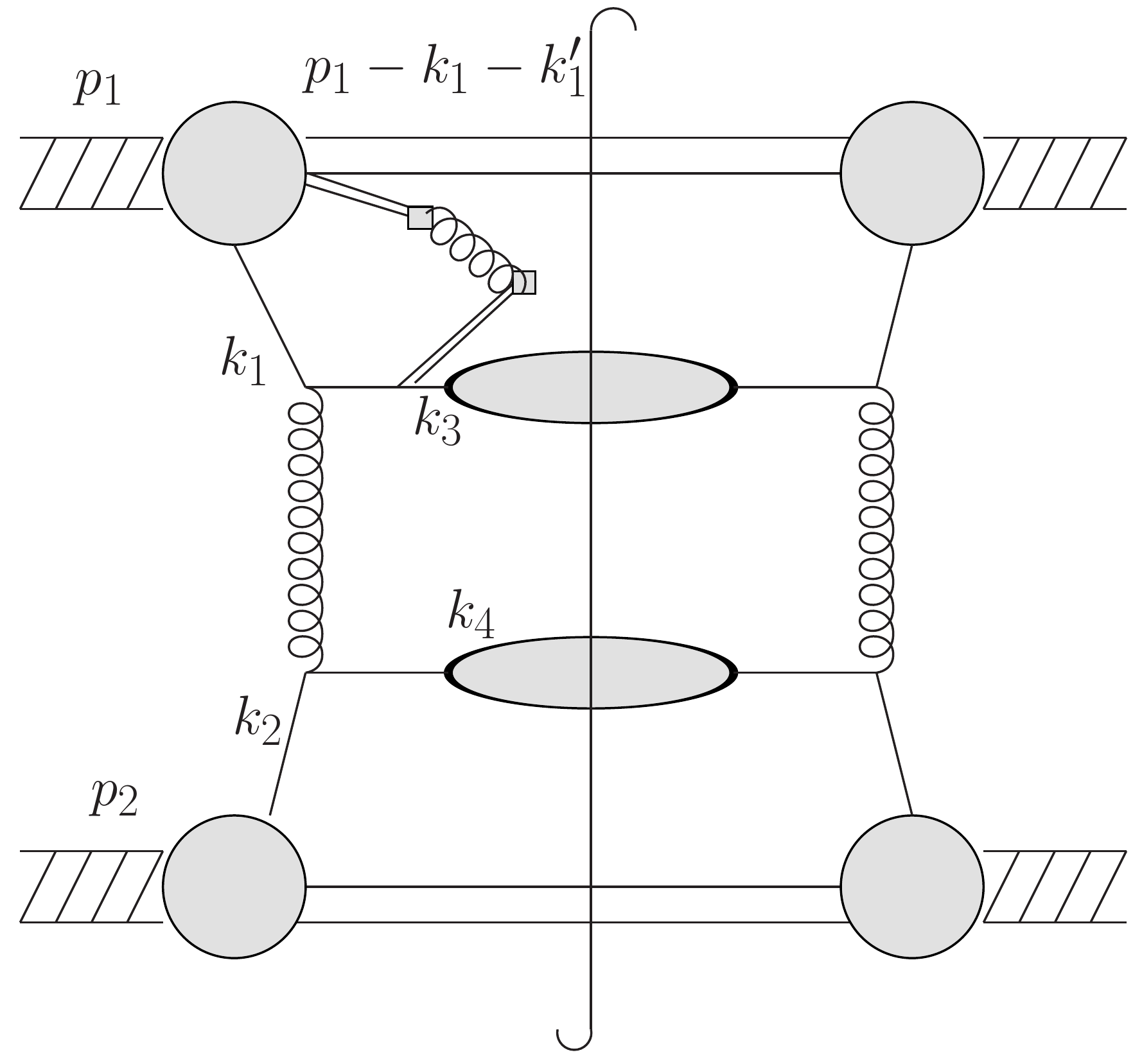}
\caption{The $S_{1a;\rho b}^{i_{1}i_{3}}$ part in (\ref{decomposition}) where the double-lines with solid box denote eikonal lines along the directions of the collinear particles.}
\label{soft1}
\end{figure*}

The $S_{1a;\rho b}^{i_{1}i_{3}}$ part can be represented by the diagram in Fig.\ref{soft1},
although the integral path of $l^{-}$ is not along the real axis at this step,
where the double-lines with solid box denote eikonal lines along the directions of the collinear particles.

For the $C_{1a;\rho b}^{i_{1}i_{3}}(l)$ part in (\ref{decomposition}), contributions of soft region  are power suppressed. We can thus take the approximation:
\begin{eqnarray}
&&\int \frac{\ud l^{+}}{2\pi}\int\frac{\ud^{2}\vec{l}_{\perp}}{(2\pi)^{2}}
\int_{C(l^{+},l_{\perp})}\frac{\ud l^{-}}{2\pi}C_{1a;\rho b}^{i_{1}i_{3}}(l)
\nonumber\\
&\simeq&
\int \frac{\ud l^{+}}{2\pi}\int\frac{\ud^{2}\vec{l}_{\perp}}{(2\pi)^{2}}
\int_{C(l^{+},l_{\perp})}\frac{\ud l^{-}}{2\pi}C_{f}t^{b}
\nonumber\\
&&
\left \{\left(\gamma^{+}
\frac{(k_{1}^{\prime +}+l^{+})\gamma^{-}}{(k_{1}^{\prime}+l)^{2}+i\epsilon}
\right)_{i_{1}j_{1}}
\Gamma_{H}^{j_{1}j_{3}}
\left(\frac{(k_{1}-l^{+})\gamma^{-}}{(k_{1}-l)^{2}+i\epsilon}\gamma^{\rho}
\right)_{i_{3}j_{3}}
\right.
\nonumber\\
&&
\left.
-\left(\gamma^{+}\frac{\gamma^{-}}{2l^{-}+i\epsilon}\right)_{i_{1}j_{1}}
\Gamma_{H}^{j_{1}j_{3}}
\left(
\frac{k_{1}^{+}\gamma^{-}}{k_{1}^{2}-2k_{1}^{+}l^{-}+i\epsilon}\gamma^{\rho}\right)_{i_{3}j_{3}}
\right \}
\nonumber\\
&&
\frac{1}{-l^{+}+i\epsilon}
\frac{1}{l^{2}+i\epsilon}
\end{eqnarray}
We notice that:
\begin{eqnarray}
&&\int \frac{\ud l^{+}}{2\pi}\int\frac{\ud^{2}\vec{l}_{\perp}}{(2\pi)^{2}}
\int_{C(l^{+},l_{\perp})}\frac{\ud l^{-}}{2\pi} (2\pi)\delta(l^{+})C_{f}t^{b}
\nonumber\\
&&
\left \{\left(\gamma^{+}
\frac{(k_{1}^{\prime +}+l^{+})\gamma^{-}}{(k_{1}^{\prime}+l)^{2}+i\epsilon}
\right)_{i_{1}j_{1}}
\Gamma_{H}^{j_{1}j_{3}}
\left(\frac{(k_{1}-l^{+})\gamma^{-}}{(k_{1}-l)^{2}+i\epsilon}\gamma^{\rho}
\right)_{i_{3}j_{3}}
\right.
\nonumber\\
&&
\left.
-\left(\gamma^{+}\frac{\gamma^{-}}{2l^{-}+i\epsilon}\right)_{i_{1}j_{1}}
\Gamma_{H}^{j_{1}j_{3}}
\left(
\frac{k_{1}^{+}\gamma^{-}}{k_{1}^{2}-2k_{1}^{+}l^{-}+i\epsilon}\gamma^{\rho}\right)_{i_{3}j_{3}}
\right \}
\frac{1}{l^{2}+i\epsilon}
\nonumber\\
&\simeq&\int\frac{\ud^{2}\vec{l}_{\perp}}{(2\pi)^{2}}
\int_{C(l_{\perp})}\frac{\ud l^{-}}{2\pi} C_{f}t^{b}
\nonumber\\
&&
\left \{\left(\gamma^{+}
\frac{k_{1}^{\prime +}\gamma^{-}}
{2k_{1}^{\prime+}l^{-}-2\vec{k_{1}^{\prime}}_{\perp}\cdot\vec{l}_{\perp}-|\vec{l}_{\perp}|^{2}
+i\epsilon}
\right)_{i_{1}j_{1}}
\Gamma_{H}^{j_{1}j_{3}}\right.
\nonumber\\
&&
\left(\frac{k_{1}\gamma^{-}}
{k_{1}^{2}-2k_{1}l^{-}+2\vec{k_{1}}_{\perp}\cdot\vec{l}_{\perp}-|\vec{l}_{\perp}|^{2}
+i\epsilon}
\gamma^{\rho}
\right)_{i_{3}j_{3}}
\nonumber\\
&&
\left.
-\left(\gamma^{+}\frac{\gamma^{-}}{2l^{-}+i\epsilon}\right)_{i_{1}j_{1}}
\Gamma_{H}^{j_{1}j_{3}}
\left(
\frac{k_{1}^{+}\gamma^{-}}{k_{1}^{2}-2k_{1}^{+}l^{-}+i\epsilon}\gamma^{\rho}\right)_{i_{3}j_{3}}
\right \}
\frac{1}{-|\vec{l}_{\perp}|^{2}+i\epsilon}
\nonumber\\
&\sim& O(\Lambda_{QCD}/Q)
\end{eqnarray}
where $C(l_{\perp})=C(0,l_{\perp})$ denote the integral path of $l^{-}$ with radius of order $|\vec{l}_{\perp}|$. We then have:
\begin{eqnarray}
\label{Mc}
&&\int \frac{\ud l^{+}}{2\pi}\int\frac{\ud^{2}\vec{l}_{\perp}}{(2\pi)^{2}}
\int_{C(l^{+},l_{\perp})}\frac{\ud l^{-}}{2\pi}C_{1a;\rho b}^{i_{1}i_{3}}(l)
\nonumber\\
&\simeq&
\int \frac{\ud l^{+}}{2\pi}\int\frac{\ud^{2}\vec{l}_{\perp}}{(2\pi)^{2}}
\int_{C(l^{+},l_{\perp})}\frac{\ud l^{-}}{2\pi}C_{f}t^{b}
\nonumber\\
&&
\left \{\left(\gamma^{+}
\frac{(k_{1}^{\prime +}+l^{+})\gamma^{-}}{(k_{1}^{\prime}+l)^{2}+i\epsilon}
\right)_{i_{1}j_{1}}
\Gamma_{H}^{j_{1}j_{3}}
\left(\frac{(k_{1}-l^{+})\gamma^{-}}{(k_{1}-l)^{2}+i\epsilon}\gamma^{\rho}
\right)_{i_{3}j_{3}}
\right.
\nonumber\\
&&
\left.
-\left(\gamma^{+}\frac{\gamma^{-}}{2l^{-}+i\epsilon}\right)_{i_{1}j_{1}}
\Gamma_{H}^{j_{1}j_{3}}
\left(
\frac{k_{1}^{+}\gamma^{-}}{k_{1}^{2}-2k_{1}^{+}l^{-}+i\epsilon}\gamma^{\rho}\right)_{i_{3}j_{3}}
\right \}
\nonumber\\
&&
P\left(\frac{1}{-l^{+}}\right)
\frac{1}{l^{2}+i\epsilon}
\end{eqnarray}
where $P$ denote the principal value.

For the second and third diagrams of Fig.\ref{subtraction}, we can repeat the similar calculations. The difference is that  coupling between soft gluon and $k_{2}$ is absorbed into past pointing Wilson line in the $S_{1c;\rho b}^{i_{1}i_{3}}(l)$. While such couplings are absorbed into future pointing Wilson lines in $S_{1a;\rho b}^{i_{1}i_{3}}(l)$ and $S_{1b;\rho b}^{i_{1}i_{3}}(l)$. This do not affect the $C_{1a;\rho b}^{i_{1}i_{3}}(l)$, $C_{1b;\rho b}^{i_{1}i_{3}}(l)$ and $C_{1c;\rho b}^{i_{1}i_{3}}(l)$ parts as there the singular point $l^{+}=0$ do not contribute to these parts.

For the fourth diagram of Fig.\ref{subtraction}, $l$ can not locate in collinear region at leading order. We deform the integral path  $l^{-}$ to lower half plane and that of $l^{+}$ to upper half plane with radius of order $|\vec{l}_{\perp}|$ to avoid the Glauber region. There are not $C_{1d;\rho b}^{i_{1}i_{3}}(l)$ term at leading order.

\begin{figure*}
\centering
\begin{tabular}{c@{\hspace*{5mm}}c}
\includegraphics[scale=0.3]{soft.pdf}
&
\includegraphics[scale=0.3]{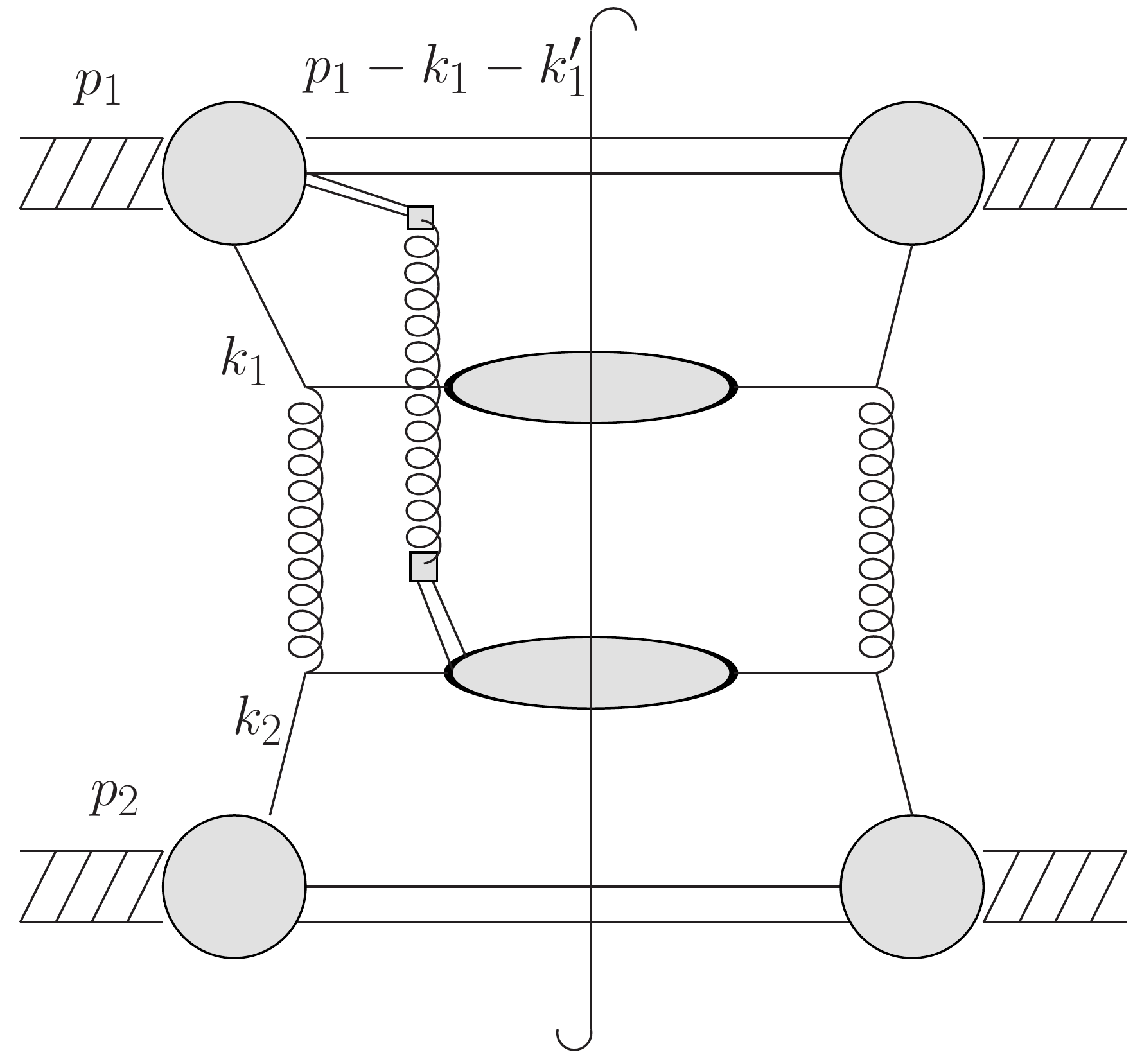}
\\
(a)&(b)
\\
\includegraphics[scale=0.3]{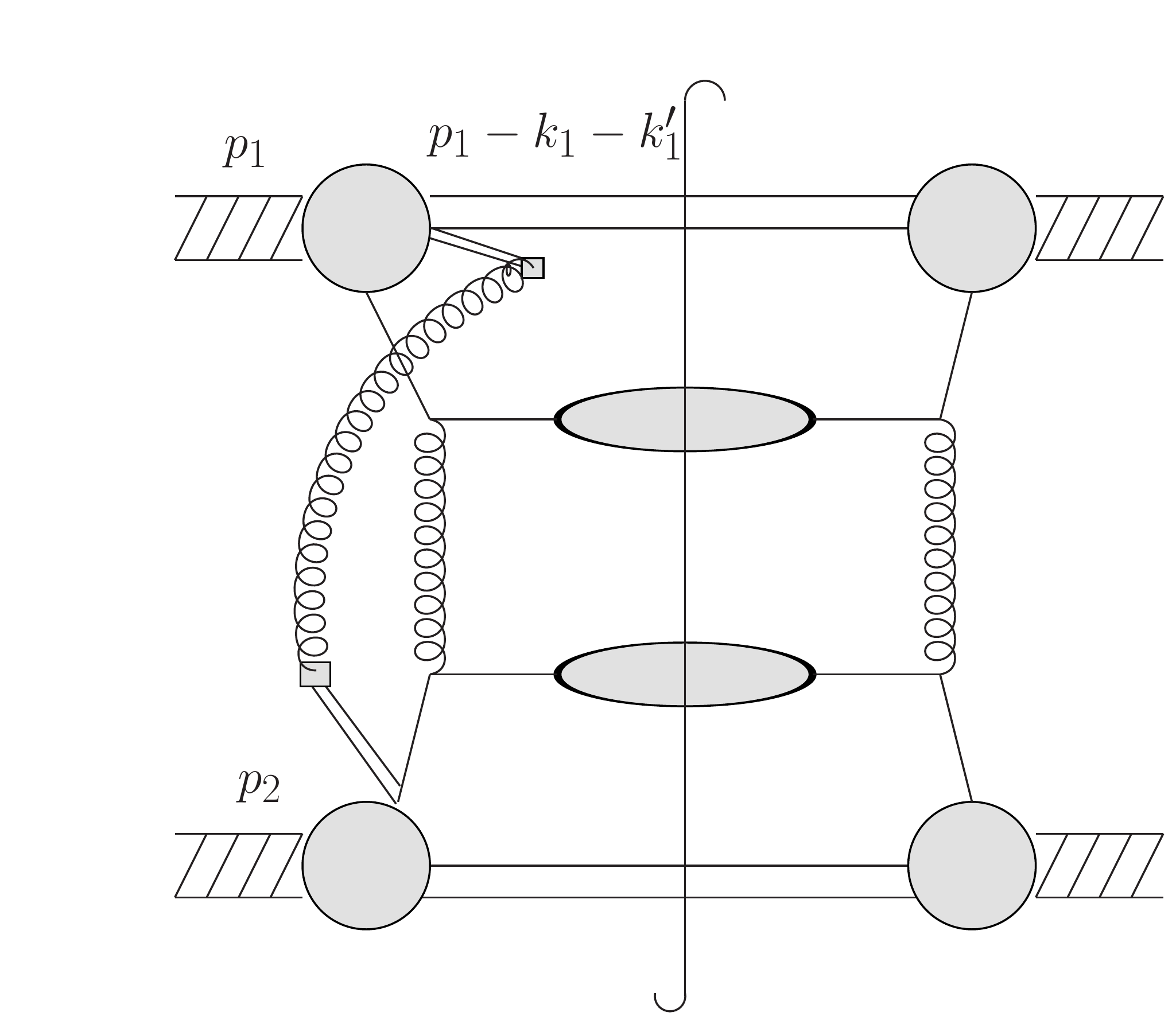}
&
\includegraphics[scale=0.3]{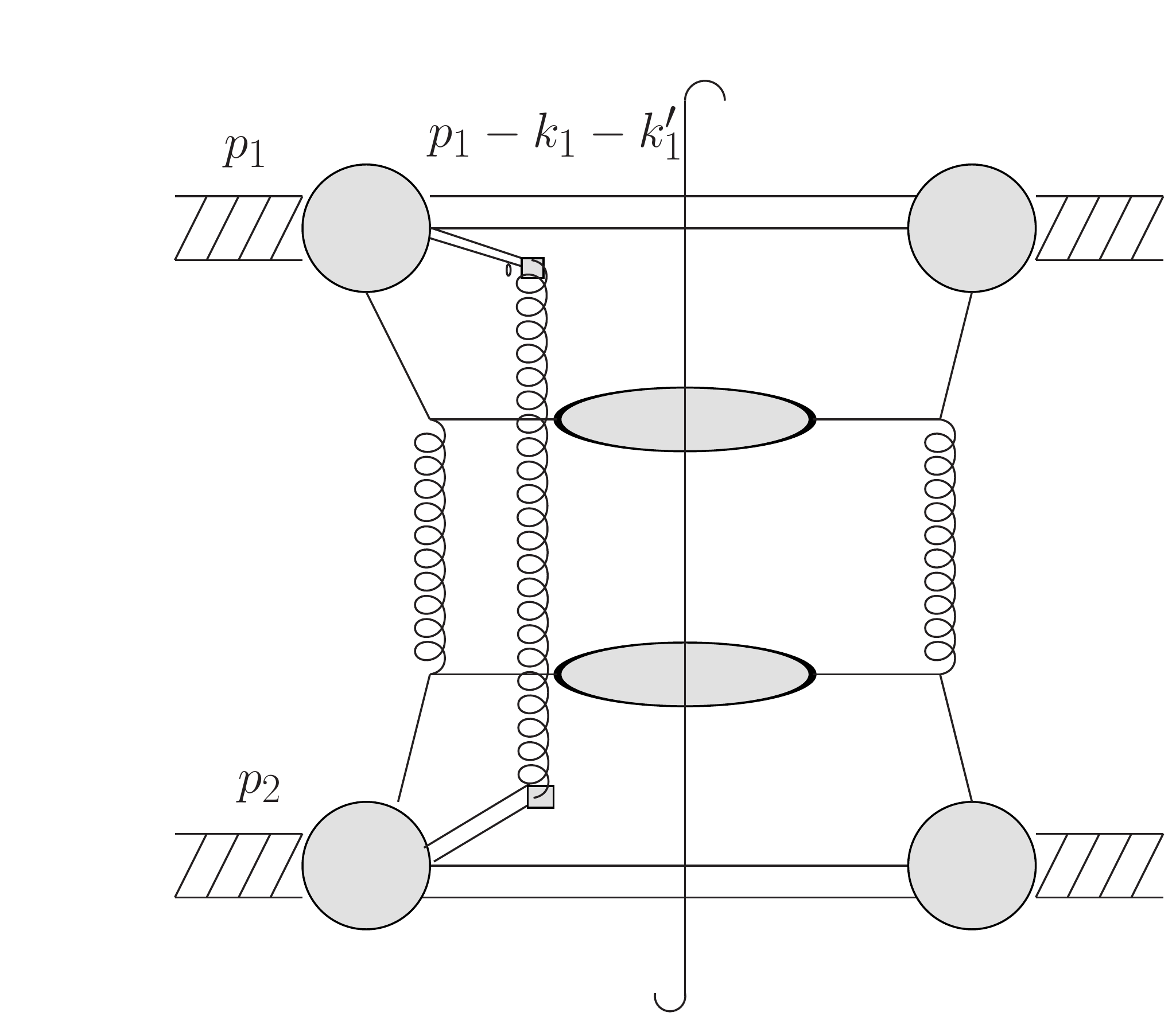}
\\
(c)&(d)
\end{tabular}
\caption{Diagrams corresponding to $S_{1a;\rho b}^{i_{1}i_{3}}(l)$, $S_{1b;\rho b}^{i_{1}i_{3}}(l)$, $S_{1c;\rho b}^{i_{1}i_{3}}(l)$ and $S_{1d;\rho b}^{i_{1}i_{3}}(l)$ terms, where the double-lines with solid box denote eikonal lines along the directions of the collinear particles. }
\label{soft}
\end{figure*}

We display the $S_{1a;\rho b}^{i_{1}i_{3}}(l)$, $S_{1b;\rho b}^{i_{1}i_{3}}(l)$, $S_{1c;\rho b}^{i_{1}i_{3}}(l)$ and $S_{1d;\rho b}^{i_{1}i_{3}}(l)$ terms in Fig.(\ref{soft})
, where the double-lines with solid box denote eikonal lines along the directions of the collinear particles.
.

\begin{figure*}
\centering
\begin{tabular}{c@{\hspace*{5mm}}c}
\includegraphics[scale=0.3]{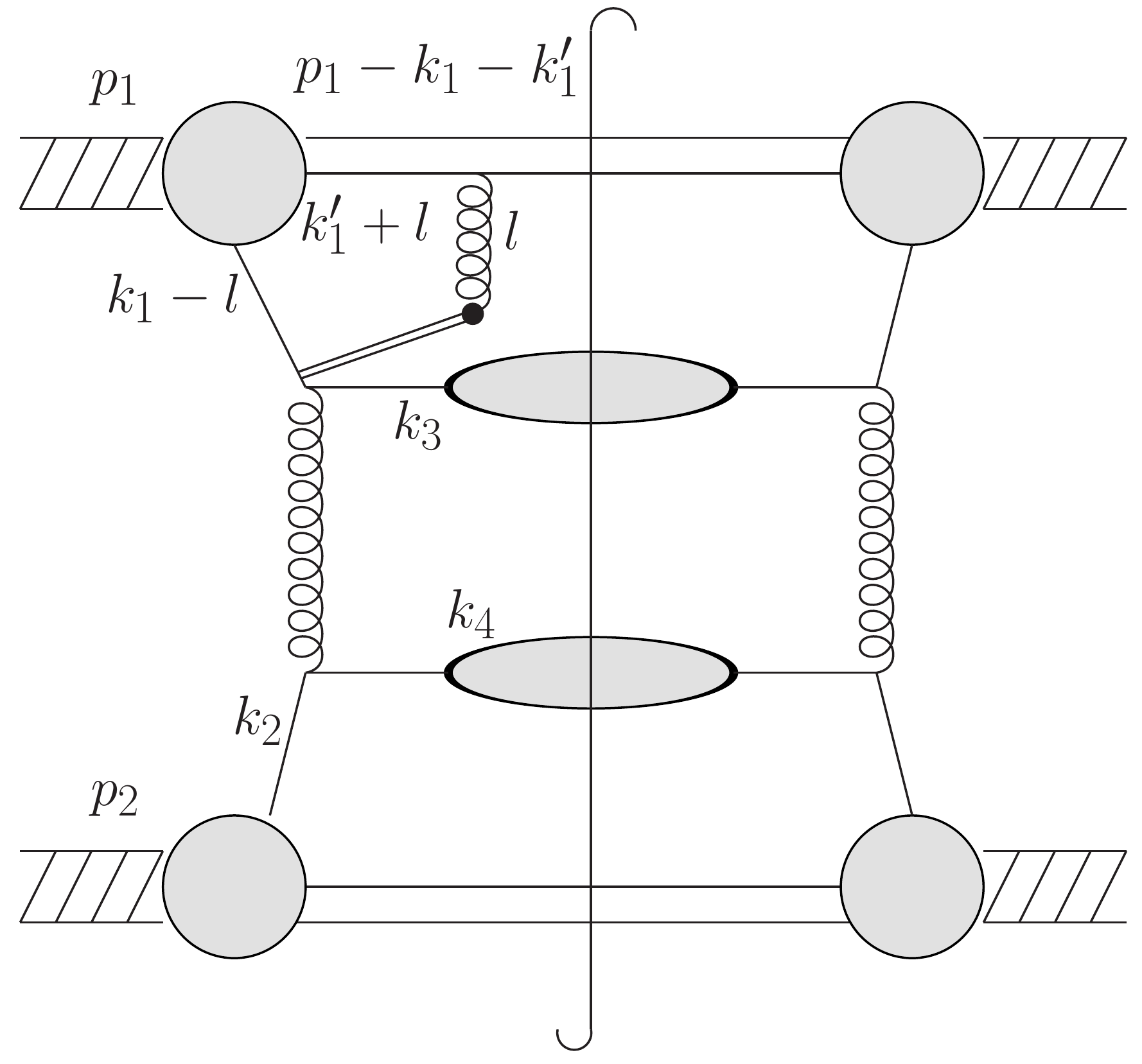}
&
\includegraphics[scale=0.3]{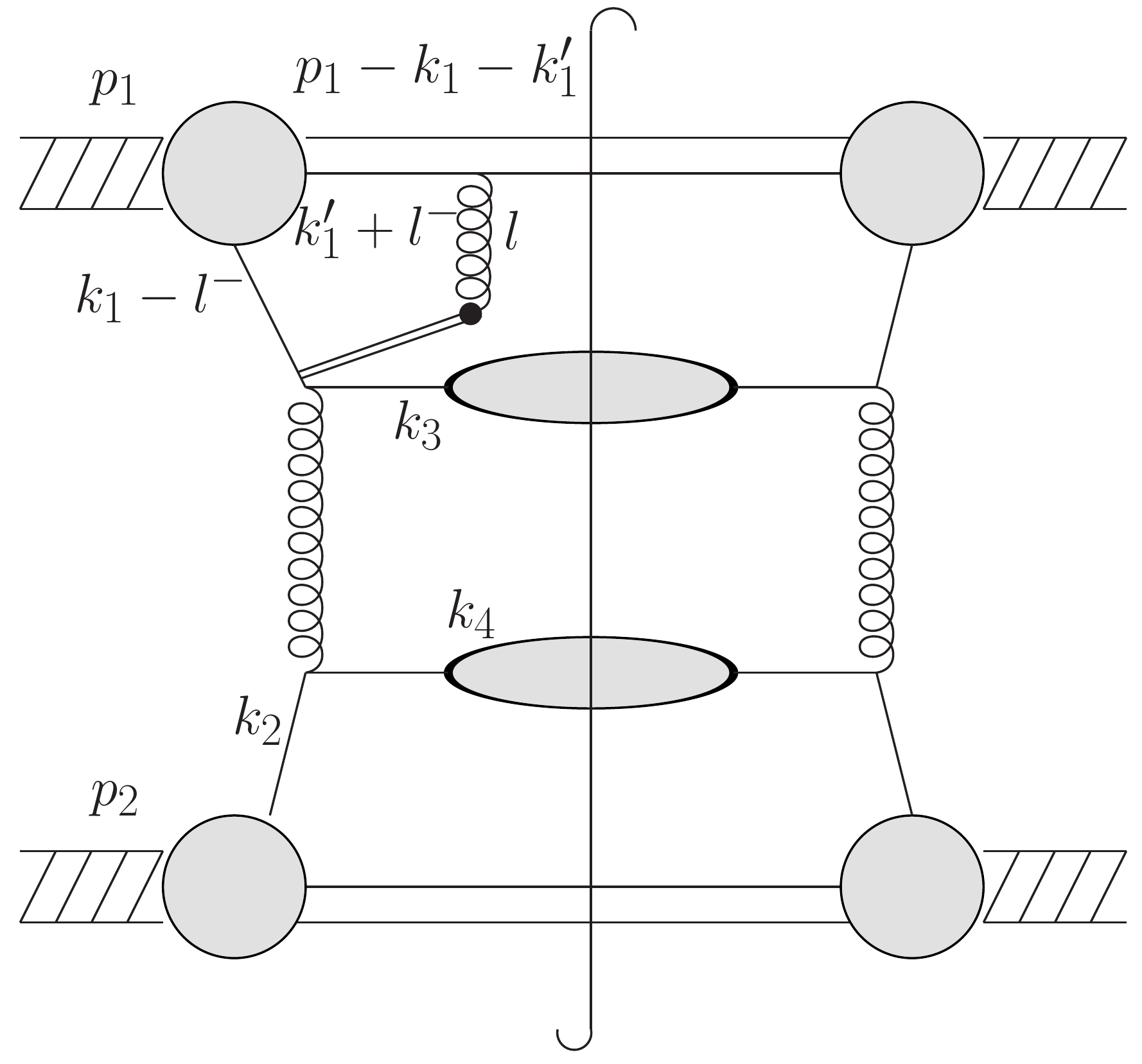}
\\
(a)&(b)
\end{tabular}
\caption{Diagrams corresponding to the summation of  $C_{1a;\rho b}^{i_{1}i_{3}}(l)$, $C_{1b;\rho b}^{i_{1}i_{3}}(l)$ and $C_{1c;\rho b}^{i_{1}i_{3}}(l)$ terms  after the summation over attachments of the collinear gluons to the hard part is perfumed, where the double-lines with solid circle represent eikonal lines along the minus direction.}
\label{collinear diagram}
\end{figure*}

For the $C_{1a;\rho b}^{i_{1}i_{3}}(l)$, $C_{1b;\rho b}^{i_{1}i_{3}}(l)$ and $C_{1c;\rho b}^{i_{1}i_{3}}(l)$ terms, there are not contributions of the singular points $l^{+}=0$. We then apply the Ward identity to the summation of these three terms. The result is the difference between the two diagrams in Fig.(\ref{collinear diagram}), where the double-lines with solid circle represent eikonal lines along the minus direction.

We then deform the integral path of $l^{-}$ to the real axis. Contributions of poles confronted in such deformation cancel out according to the similar calculations we performed to deform the integral path initially. Fig.\ref{soft} and Fig.\ref{collinear diagram} take the form we demanded. It is,  however,  not easy to extend such calculations to higher order. Instead of explicit calculations, we consider this problem in the frame of effective theory in next section.
\\

\section{Cancellation of Pinch Singularities in Glauber Region in Effective Theory}
\label{deformation}

In this section, we prove that one can deform the integral path so that the Grammer-Yennie approximation works in couplings between soft gluons and collinear particles in processes considered in this paper. We denote the momentum of a soft gluon as $q^{\mu}$, which is defined as flow into collinear particles. If the soft gluon $q$ couple to particles collinear to one initial hadron, then we can deform the integral path of $n\cdot q$ to upper half plane so that the the Grammer-Yennie approximation works, where $n^{\mu}$ denotes the direction of the collinear particles. If the soft gluon $q$ couple to particles collinear to other directions, then we can deform the integral path of $n\cdot q$ to lower half plane so that the the Grammer-Yennie approximation works.

The hard subprocess between partons is nearly local in coordinate space with uncertainty of order $1/Q$. We denote the space time region in which the hard collision occur as $H(x,1/Q)$, where $x$ is the center of the hard subprocess. After or before the hard collision, there are interactions between collinear and soft particles  which are responsible for the formation of final hadrons or the production of initial active partons. The space-time uncertainty of such interactions are of order $1/\Lambda_{QCD}$. We denote the space-time region of such interactions as $S(x,1/\Lambda_{QCD})$.    We also denote the remaining space-time region as $C(x)$. Different jets and soft hadrons separate from each other in the region $C(x)$. We can neglect interactions between different jets and soft hadrons in the region $C(x)$ as they are color singlets and sperate from each other far enough.

As in \cite{zhou:2013dr}, We fist consider the classical configuration of quark and gluon fields in the region $S(x,1/\Lambda_{QCD})$. Different jets separate from each other in this region although they exchange soft gluons. We denote the region which the jet collinear to $n^{\mu}$ locate in as $y_{n}$. We also denote the region in which there are not collinear jets as $y_{s}$.
($y_{n}\subset S(x,1/\Lambda_{QCD})$, $y_{s}\subset S(x,1/\Lambda_{QCD})$). We define the soft fields according to the manner:
\begin{equation}
(D_{s\mu}G_{s}^{\mu\nu})^{a}(y_{n})=g\bar{\psi}_{s}\gamma^{\mu}t^{a}\psi_{s}(y_{n})
,\quad
A_{s}^{\mu}(y_{s})\equiv A^{\mu}(y_{s})
\end{equation}
and
\begin{equation}
\not\!D_{s}\psi_{s}(y_{n})=0,\quad \psi_{s}(y_{s})\equiv\psi(y_{s})
\end{equation}
where
\begin{equation}
D_{s}^{\mu}\equiv\partial^{\mu}-ig A_{s}^{\mu},\quad G_{s}^{\mu\nu}\equiv\frac{i}{g}[D_{s}^{\mu},D_{s}^{\nu}]
\end{equation}
The equation should be solved perturbatively in the region $y_{n}$, that is, soft fields in the region $y_{n}$ is defined according to perturbation theory.
We then define the collinear fields:
\begin{equation}
\psi_{n}(y_{n})\equiv\psi(y_{n})-\psi_{s}(y_{n})
\end{equation}
\begin{equation}
\psi_{n}(y_{m})=\psi_{n}(y_{s})=0 (m^{\mu}\neq n^{\mu})
\end{equation}
\begin{equation}
A_{n}^{\mu}(y_{n})\equiv A^{\mu}(y_{n})-A_{s}^{\mu}(y_{n})
\end{equation}
\begin{equation}
 A_{n}^{\mu}(y_{m})=A_{n}^{\mu}(y_{m})=0 (m^{\mu}\neq n^{\mu})
\end{equation}

We then write the classical Lagrangian density of QCD as:
\begin{eqnarray}
\mathcal{L}&=&\sum_{n^{\mu}}\mathcal{L}_{n}
+\mathcal{L}_{s}
\nonumber\\
&=&\sum_{n^{\mu}}i\bar{\psi}_{n}(\not\!\partial-ig\not\!A_{n}-ig\not\!A_{s})\psi_{n}
\nonumber\\
&&
+\frac{1}{2g^{2}}\sum_{n^{\mu}}tr_{c}\{[\partial^{\mu}-igA_{n}^{\mu}-igA_{s}^{\mu},
\nonumber\\
&&
\partial^{\nu}-igA_{n}^{\nu}-igA_{s}^{\nu}]^{2}
\nonumber\\
&&-
[\partial^{\mu}-igA_{s}^{\mu},
\partial^{\nu}-igA_{s}^{\nu}]^{2}\}
\nonumber\\
&&+i\bar{\psi}_{s}(\not\!\partial-ig\not\!A_{s})\psi_{s}
\nonumber\\
&&
+\frac{1}{2g^{2}}tr_{c}\{[\partial^{\mu}-igA_{s}^{\mu},
\partial^{\nu}-igA_{s}^{\nu}]^{2}\}
\end{eqnarray}
where we have dropped the coupling between soft fermion fields and collinear fields as they are power suppressed(\cite{S:1978,zhou:2013dr}).

We decompose the region $S(x,1/\Lambda_{QCD})$ according to the manner:
\begin{eqnarray}
S(x,1/\Lambda_{QCD})&=&S_{s}(x,1/\Lambda_{QCD})+\sum_{n^{\mu}\neq n_{1}^{\mu},n^{\mu}\neq n_{2}^{\mu}}S_{n}(x,1/\Lambda_{QCD})
\nonumber\\
&&
+S_{n_{1}}(x,1/\Lambda_{QCD})+S_{n_{2}}(x,1/\Lambda_{QCD})
\end{eqnarray}
where $n_{1}^{\mu}$ and $n_{2}^{\mu}$ are defended in (\ref{n+}), $S_{n}(x,1/\Lambda_{QCD})$ denotes the region which the jet collinear to $n^{\mu}$ locate in, $S_{s}(x,1/\Lambda_{QCD})$ denotes the region in which there are not collinear jets.

There are not jets collinear to $n^{\mu}$($n^{\mu}\neq n_{1}^{\mu},n^{\mu}\neq n_{2}^{\mu}$) before the hard collision. We can thus deform the integral path(\cite{S:1978,CS:1981S}) so that the Grammer-Yennie approximation works in the coupling between soft fields and collinear fields in the region
$S_{n}(x)$($n^{\mu}\neq n_{1}^{\mu},n^{\mu}\neq n_{2}^{\mu}$). We redefine the effective fields(\cite{SCET:2001,SCET:2002,BFPRS:2002,zhou:2013dr}):
\begin{equation}
\psi_{n}^{(0)}(y_{n})=Y_{n}^{\dag}\psi_{n}(y_{n})
,\quad A_{n}^{(0)\mu}(y_{n})=Y_{n}^{\dag}A_{n}^{\mu}(y_{n})Y_{n}(y_{n})
\end{equation}
where $n^{\mu}\neq n_{1}^{\mu},n^{\mu}\neq n_{2}^{\mu}$ and
\begin{equation}
\label{soft Wilson line}
Y_{n}(y_{n})=(P\exp(ig\int_{0}^{\infty}\ud s n\cdot A_{s}(y_{n}+sn))^{\dag}
\end{equation}
The Wilson line travel from $y_{n}$ to $\infty$ according to the deformation of the integral path of soft gluons.(\cite{S:1978,CS:1981S}) Coupling between the classical fides $\psi_{n}^{(0)}$($A_{n}^{(0)\mu}$) and the classical fields $A_{s}$ do not contribute to the process at leading order in $\Lambda_{QCD}/Q$. We can then write the effective classical Lagrangian density in the region $S(x,1/\Lambda_{QCD})$ as:
\begin{equation}
\mathcal{L}_{\Lambda}=\sum_{n^{\mu}\neq n_{1}^{\mu},n^{\mu}\neq n_{2}^{\mu}}\mathcal{L}_{n}^{(0)}+\mathcal{L}_{n_{1}}
+\mathcal{L}_{n_{2}}+\mathcal{L}_{s}
\end{equation}
\begin{eqnarray}
\mathcal{L}_{n}^{(0)}&=&i\bar{\psi}_{n}^{(0)}(\widetilde{\not\!\partial}_{n}
-ig\not\!A_{n}^{(0)})\psi_{n}^{(0)}
\nonumber\\
      &&+\frac{1}{2g^{2}}
      tr\left\{([\widetilde{\partial}_{n}^{\phantom{n}\mu}-igA_{n}^{(0)\mu},
        \widetilde{\partial}_{n}^{\phantom{n}\nu}-igA_{n}^{(0)\nu}])^{2}\right\}
\end{eqnarray}
\begin{eqnarray}
\mathcal{L}_{n_{1}}&=&i\bar{\psi}_{n_{1}}(\not\!\partial-ig\not\!A_{n_{1}}-ig\not\!A_{s})\psi_{n_{1}}
\nonumber\\
&&
+\frac{1}{2g^{2}}tr_{c}\left\{[\partial^{\mu}-igA_{n_{1}}^{\mu}-igA_{s}^{\mu},
\partial^{\nu}-igA_{n_{1}}^{\nu}-igA_{s}^{\nu}]^{2}
\right.
\nonumber\\
&&
\left.
[\partial^{\mu}-igA_{s}^{\mu},
\partial^{\nu}-igA_{s}^{\nu}]^{2}
\right
\}
\end{eqnarray}
\begin{eqnarray}
\mathcal{L}_{n_{2}}&=&
i\bar{\psi}_{n_{2}}(\not\!\partial-ig\not\!A_{n_{2}}-ig\not\!A_{s})\psi_{n_{2}}
\nonumber\\
&&
+\frac{1}{2g^{2}}tr_{c}\left\{[\partial^{\mu}-igA_{n_{2}}^{\mu}-igA_{s}^{\mu},
\partial^{\nu}-igA_{n_{2}}^{\nu}-igA_{s}^{\nu}]^{2}
\right.
\nonumber\\
&&
\left.
[\partial^{\mu}-igA_{s}^{\mu},
\partial^{\nu}-igA_{s}^{\nu}]^{2}
\right
\}
\end{eqnarray}
\begin{eqnarray}
\mathcal{L}_{s}&=&i\bar{\psi}_{s}(\not\!\partial-ig\not\!A_{s})\psi_{s}
\nonumber\\
&&
+\frac{1}{2g^{2}}tr_{c}\{[\partial^{\mu}-igA_{s}^{\mu},
\partial^{\nu}-igA_{s}^{\nu}]^{2}\}
\end{eqnarray}
where the derivative $\widetilde{\partial}_{n}^{\phantom{\mu}}$ is defined as:
\begin{equation}
\widetilde{\partial}_{n}^{\phantom{n}\mu}\psi_{n}^{(0)}\equiv\partial^{\mu}\psi_{n}^{(0)}, \quad
\widetilde{\partial}_{n}^{\phantom{n}\mu}A_{n}^{(0)\nu}\equiv\partial^{\mu}A_{n}^{(0)\nu}
\end{equation}
\begin{equation}
\widetilde{\partial}_{n}^{\phantom{n}\mu} A_{s}^{\nu}\equiv\bar{n}^{\mu}n\cdot\partial  A_{s}^{\nu}
\end{equation}

We can then quantize the Lagrangian density $\mathcal{L}_{\Lambda}$ by quantize the fields $\psi_{n}^{(0)}$, $A_{n}^{(0)}$($n^{\mu}\neq n_{1}^{\mu},n^{\mu}\neq n_{2}^{\mu}$),  $\psi_{s}$,  $A_{s}$, $\psi_{n_{i}}$ and $A_{n_{i}}$($i=1,2$) independently. While dealing with couplings  between soft particles and collinear particles ($n^{\mu}\neq n_{1}^{\mu},n^{\mu}\neq n_{2}^{\mu}$) in the process considered in this paper, such quantization scheme gives the same result as that in QCD at leading order in $\Lambda_{QCD}/Q$(\cite{zhou:2013dr}). These fields should be quantized as fulfil the boundary condition provided by classical configurations of effective fields in the region $C(x)$ and the conditions:
\begin{equation}
\psi_{n_{i}}(y_{m})=\psi_{n_{i}}(y_{s})=0 (m^{\mu}\neq n_{i}^{\mu})
\end{equation}
\begin{equation}
 A_{n_{i}}^{\mu}(y_{m})=A_{n_{i}}^{\mu}(y_{m})=0 (m^{\mu}\neq n_{i}^{\mu})
\end{equation}
for $i=1,2$ and
\begin{equation}
\psi_{n}^{(0)}(y_{m})=\psi_{n}^{(0)}(y_{s})=0 (m^{\mu}\neq n^{\mu})
\end{equation}
\begin{equation}
 A_{n}^{(0)\mu}(y_{m})=A_{n}^{(0)\mu}(y_{m})=0 (m^{\mu}\neq n^{\mu})
\end{equation}
for $n^{\mu}\neq n_{1}^{\mu},n^{\mu}\neq n_{2}^{\mu}$

Diagrams with disconnected hard parts do not contribute to the process at leading order in $\Lambda_{QCD}/Q$. (\cite{P:1980,PT:1982,LS:1985}). We bring in an effective hard vertex $\Pi(x)$ to describe the effects of hard sub-process. According to the gauge invariance, we have:
\begin{eqnarray}
\Pi(x)&=&\Pi(Y_{n}\psi_{n}^{(0)}(x+n\cdot y_{n}\bar{n}),Y_{n}A_{n}^{(0)}Y_{n}^{\dag}(x+n\cdot y_{n}^{\prime}\bar{n} ),\ldots,
\nonumber\\
&&
(\psi_{n_{1}}+\psi_{s})(x+n_{1}\cdot y_{n_{1}}\bar{n_{1}}),
(A_{n_{1}}+A_{s})(x+n_{1}\cdot y_{n_{1}}^{\prime}\bar{n_{1}}),
\nonumber\\
&&
(\psi_{n_{2}}+\psi_{s})(x+n_{2}\cdot y_{n_{2}}\bar{n_{2}}),
(A_{n_{2}}+A_{s})(x+n_{2}\cdot y_{n_{2}}^{\prime}\bar{n_{2}}))
\end{eqnarray}
where we have dropped the $\psi_{s}(x+y_{s})$ and $A_{s}(x+y_{s})$ terms as couplings between soft particles and modes with $|k^{2}|\gtrsim Q^{2}$ are suppressed as power of $\Lambda_{QCD}/Q$.(\cite{S:1978}). $\Pi(x)$ is the hard vertex, we have:
\begin{equation}
\label{locality}
y_{n}^{0}\sim y_{n}^{\prime 0}\sim 1/Q
\end{equation}
To get $\Pi(x)$, one should integrate out degrees of freedom with $|k^{2}|\gtrsim Q^{2}$. We assume that this can be performed without breaking the gauge invariance.

 According to our quantization scheme, time evolution of the effective fields are:
\begin{equation}
Y_{n}\psi_{n}^{(0)}(x+y_{n})=
e^{iH_{\Lambda}x^{0}}Y_{n}\psi_{n}^{(0)}(\vec{x}+y_{n})e^{-iH_{\Lambda}x^{0}}
\end{equation}
\begin{equation}
Y_{n}A_{n}^{(0)}Y_{n}^{\dag}(x+y_{n}^{\prime})=
e^{iH_{\Lambda}x^{0}}Y_{n}A_{n}^{(0)}Y_{n}^{\dag}(x+y_{n}^{\prime})e^{-iH_{\Lambda}x^{0}}
\end{equation}
\begin{equation}
(\psi_{n_{1}}+\psi_{s})(x+y_{n_{1}})=
e^{iH_{\Lambda}x^{0}}(\psi_{n_{1}}+\psi_{s})(\vec{x}+y_{n_{1}})e^{-iH_{\Lambda}x^{0}}
\end{equation}
\begin{equation}
(A_{n_{1}}+A_{s})(x+y_{n_{1}}^{\prime})=
e^{iH_{\Lambda}x^{0}}(A_{n_{1}}+A_{s})(\vec{x}+y_{n_{1}}^{\prime})e^{-iH_{\Lambda}x^{0}}
\end{equation}
\begin{equation}
(\psi_{n_{2}}+\psi_{s})(x+y_{n_{2}})=
e^{iH_{\Lambda}x^{0}}(\psi_{n_{2}}+\psi_{s})(\vec{x}+y_{n_{2}})e^{-iH_{\Lambda}x^{0}}
\end{equation}
\begin{equation}
(A_{n_{2}}+A_{s})(x+y_{n_{2}}^{\prime})=
e^{iH_{\Lambda}x^{0}}(A_{n_{2}}+A_{s})(\vec{x}+y_{n_{2}}^{\prime})e^{-iH_{\Lambda}x^{0}}
\end{equation}
where $H_{\Lambda}$ is the Hamiltonian corresponding to $\mathcal{L}_{\Lambda}$. We then have:
\begin{eqnarray}
\Pi(x)&=&e^{iH_{\Lambda}x^{0}}\Pi(\vec{x})e^{-iH_{\Lambda}x^{0}}
\end{eqnarray}

We then write the transition amplitude of the process in the frame of the effective theory:
\begin{equation}
\lim_{T\to\infty}\quad _{T}\big<H_{3}H_{4}X|e^{-iH_{\Lambda}(T-x^{0})}
\Pi(\vec{x})e^{-iH_{\Lambda}(x^{0}+T)}|p_{1}p_{2}\big>_{-T}
\end{equation}
where the subscripts $T$ and $-T$ denote the sates at the time $t=T$ and $t=-T$ in Scr$\ddot{o}$dinger picture. We take the modular square of such amplitude and make the summation over all possible states $X$ and the point at which the hard collision occur, this gives:
\begin{eqnarray}
\label{propability}
&&\lim_{T\to\infty}\sum_{X}\int\ud^{4}x_{2}\int\ud^{4}x_{1}
\nonumber\\
&&
_{-T}\big<p_{1}p_{2}|e^{iH_{\Lambda}(x_{2}^{0}+T)}
\Pi^{\dag}(\vec{x}_{1})e^{iH_{\Lambda}(T-x_{2}^{0})}|H_{3}H_{4}X\big>_{T}
\nonumber\\
&&_{T}\big<H_{3}H_{4}X|e^{-iH_{\Lambda}(T-x_{1}^{0})}
\Pi(\vec{x}_{2})e^{-iH_{\Lambda}(x_{1}^{0}+T)}|p_{1}p_{2}\big>_{-T}
\end{eqnarray}

We write $H_{\Lambda}$ as:
\begin{equation}
H_{\Lambda}=\sum_{n^{\mu}\neq n_{1}^{\mu},n^{\mu}\neq n_{2}^{\mu}}H_{n}^{(0)}+H_{n_{1}}
+H_{n_{2}}+H_{s}\equiv\sum_{n^{\mu}\neq n_{1}^{\mu},n^{\mu}\neq n_{2}^{\mu}}H_{n}^{(0)}+H_{cs}
\end{equation}
where $H_{n}^{(0)}$($n^{\mu}\neq n_{1}^{\mu},n^{\mu}\neq n_{2}^{\mu}$), $H_{n_{1}}$,
$H_{n_{2}}$ and $H_{s}$ represent the Hamiltonian corresponding to $\mathcal{L}_{n}^{(0)}$($n^{\mu}\neq n_{1}^{\mu},n^{\mu}\neq n_{2}^{\mu}$), $\mathcal{L}_{n_{1}}$,
$\mathcal{L}_{n_{2}}$ and $\mathcal{L}_{s}$ respectively.
We have $[H_{n}^{(0)},H_{cs}]=0$ and  write (\ref{propability}) as:
\begin{eqnarray}
&&\lim_{T\to\infty}\sum_{X}\int\ud^{4}x_{2}\int\ud^{4}x_{1}
\nonumber\\
&&
_{-T}\big<p_{1}p_{2}|e^{iH_{cs}(x_{2}^{0}+T)}
e^{i\sum_{n^{\mu}\neq n_{1}^{\mu},n^{\mu}\neq n_{2}^{\mu}}H_{n}^{(0)}(x_{2}^{0}+T)}
\Pi^{\dag}(\vec{x}_{2})
\nonumber\\
&&
e^{i\sum_{n^{\mu}\neq n_{1}^{\mu},n^{\mu}\neq n_{2}^{\mu}}H_{n}^{(0)}(T-x_{2}^{0})}e^{iH_{cs}(T-x_{2}^{0})}|H_{3}H_{4}X\big>_{T}
\nonumber\\
&&_{T}\big<H_{3}H_{4}X|e^{-iH_{cs}(T-x_{1}^{0})}e^{-i\sum_{n^{\mu}\neq n_{1}^{\mu},n^{\mu}\neq n_{2}^{\mu}}H_{n}^{(0)}(T-x_{1}^{0})}
\Pi(\vec{x}_{1})
\nonumber\\
&&
e^{-i\sum_{n^{\mu}\neq n_{1}^{\mu},n^{\mu}\neq n_{2}^{\mu}}H_{n}^{(0)}(x_{1}^{0}+T)}e^{-iH_{cs}(x_{1}^{0}+T)}|p_{1}p_{2}\big>_{-T}
\end{eqnarray}
The states $|H_{3}H_{4}X\big>_{T}$ form an invariant subspace of $H_{cs}$ as $H_{3}$ and $H_{4}$ decouple from $H_{cs}$. We have:
\begin{eqnarray}
&&e^{iH_{cs}(T-x_{1}^{0})}|H_{3}H_{4}X\big>_{T}
\quad_{T}\big<H_{3}H_{4}X|e^{-iH_{cs}(T-x_{2}^{0})}
\nonumber\\
&=&e^{-iH_{cs}x_{1}^{0}}|H_{3}H_{4}X\big>_{T}
\quad_{T}\big<H_{3}H_{4}X|e^{iH_{cs}x_{2}^{0}}
\nonumber\\
&=&e^{-iH_{cs}(x_{1}^{0}+T)}|H_{3}H_{4}X\big>_{T}
\quad_{T}\big<H_{3}H_{4}X|e^{iH_{cs}(x_{2}^{0}+T)}
\end{eqnarray}
We can then write (\ref{propability}) as:
\begin{eqnarray}
\label{initial evolution}
&&\lim_{T\to\infty}\sum_{X}\int\ud^{4}x_{2}\int\ud^{4}x_{1}
\nonumber\\
&&
_{-T}\big<p_{1}p_{2}|e^{iH_{cs}(x_{2}^{0}+T)}
e^{i\sum_{n^{\mu}\neq n_{1}^{\mu},n^{\mu}\neq n_{2}^{\mu}}H_{n}^{(0)}(x_{2}^{0}+T)}
\Pi^{\dag}(\vec{x}_{2})
\nonumber\\
&&
e^{i\sum_{n^{\mu}\neq n_{1}^{\mu},n^{\mu}\neq n_{2}^{\mu}}H_{n}^{(0)}(T-x_{2}^{0})}e^{-iH_{cs}(x_{2}^{0}+T)}|H_{3}H_{4}X\big>_{T}
\nonumber\\
&&_{T}\big<H_{3}H_{4}X|e^{iH_{cs}(x_{1}^{0}+T)}e^{-i\sum_{n^{\mu}\neq n_{1}^{\mu},n^{\mu}\neq n_{2}^{\mu}}H_{n}^{(0)}(T-x_{1}^{0})}
\Pi(\vec{x}_{1})
\nonumber\\
&&
e^{-i\sum_{n^{\mu}\neq n_{1}^{\mu},n^{\mu}\neq n_{2}^{\mu}}H_{n}^{(0)}(x_{1}^{0}+T)}e^{-iH_{cs}(x_{1}^{0}+T)}|p_{1}p_{2}\big>_{-T}
\end{eqnarray}
According to (\ref{locality}),
we see that contributions of coupling between remnants of  a initial hadron and other particles(soft or collinear to the initial hadron) after the hard collision cancel out at leading order.

We notice that:
\begin{eqnarray}
e^{-iH_{cs}(x_{2}^{0}-x_{1}^{0})}&=&e^{-i(H_{cs})_{0}x_{1}^{0}}(Te^{-i\int_{x_{1}^{0}}^{x_{2}^{0}}\ud y^{0}(H_{cs})_{I}(y^{0})})e^{i(H_{cs})_{0}x_{2}^{0}}
\nonumber\\
&&e^{-i(H_{cs})_{0}x_{1}^{0}}
(\sum_{n=0}^{\infty}\frac{(-i)^{n}}{n!}\int_{x_{1}^{0}}^{x_{2}^{0}}\ud y_{1}^{0}\ldots \int_{x_{1}^{0}}^{x_{2}^{0}}\ud y_{n}^{0}
\nonumber\\
&&
T[(H_{cs})_{I}(y_{1}^{0})\ldots (H_{cs})_{I}(y_{n}^{0})]
e^{i(H_{cs})_{0}x_{2}^{0}}
\end{eqnarray}
for arbitrary $x_{1}$ and $x_{2}$, where $(H_{cs})_{0}$ represents the free part of $H_{cs}$, $(H_{cs})_{I}$ represents the interaction part of $H_{cs}$ in the interaction picture. Thus the $\delta$-function  of momenta conversation at a vertex of coupling between remnants of  a initial hadron and other particles should be substituted with the function:
\begin{equation}
(2\pi)^{4}\delta^{(4)}(q_{i})\to\frac{i(2\pi)^{3}\delta^{(3)}(\vec{q}_{i})}{q_{i}^{0}+i\epsilon}e^{-iq_{1}^{0}x^{0}}
\end{equation}
in (\ref{initial evolution}), where $x^{0}=x_{1}^{0}$ or $x^{0}=x_{1}^{0}$ in (\ref{initial evolution}), $q_{i}^{0}$ denotes the total momenta flowing into the vertex.
We consider the coupling between remnants of  the initial hadron that collinear-to-plus and other particles. We make the substitution:
\begin{equation}
\frac{i(2\pi)^{3}\delta^{(3)}(\vec{q}_{i})}{q_{i}^{0}+i\epsilon}
\to \frac{\sqrt{2}i(2\pi)^{3}\delta^{(3)}(\vec{q}_{i})}{q_{i}^{-}+i\epsilon}
\end{equation}
We then take the approximation:
\begin{equation}
e^{-iq_{i}^{0}x^{0}}\simeq e^{-i\frac{1}{\sqrt{2}}q_{i}^{+}x^{0}}
\end{equation}
as we can drop the component $q_{i}^{-}$ in the $\delta$-function of momenta conversation of the whole process. For the $\delta$-function $\delta(q_{i}^{3})$, we have:
\begin{equation}
\delta(q_{i}^{3})=\sqrt{2}\delta(q_{i}^{+}-q_{i}^{-})
\end{equation}
We can then take $q_{i}^{-}$ as independent variable and integrate out $\delta(q_{i}^{3})$ to determine $q_{i}^{+}$.   In the propagators of particles collinear-to-plus with momenta $k$, we can drop the small parts of $k^{+}$. This is equivalent to take the approximation:
\begin{equation}
\delta(q_{i}^{3})\simeq \sqrt{2}\delta(q_{i}^{+})
\end{equation}
According to these analyses, we have:
\begin{equation}
\frac{i(2\pi)^{3}\delta^{(3)}(\vec{q}_{i})}{q_{i}^{0}+i\epsilon}
e^{-iq_{i}^{0}x^{0}}
\simeq \frac{2i(2\pi)\delta(q_{i}^{+})}{q_{i}^{-}+i\epsilon}
(2\pi)^{2}\delta^{(2)}(\vec{q}_{i\perp})
e^{-i\frac{1}{\sqrt{2}}q_{i}^{+}x^{0}}
\end{equation}
We notice that minus momenta of lines at the vertex are independent of each other in this case. Singular point locating in the Glauber region can only be produced by the $\frac{1}{q_{i}^{-}+i\epsilon}$ term. If there is one Glauber gluon flow into the vertex, the momentum of which we denote as $q_{i}^{G}$. Then the other end of the $q_{i}^{G}$ should be collinear particles with large minus momenta if we work in the Feynman gauge. Thus we can deform the integral path of $q_{i}^{G-}$ to the upper half plane with radius of order $\min\{|\vec{q_{i}^{G}}_{\perp}|,|\vec{q_{i}^{G}}_{\perp}|^{2}/|q_{i}^{G +}|\}$
so that the Gramma-Yennie approximation works.
We have proved our claim at the beginning of this section.
\\

\section{Wilson Lines of Collinear and Soft Gluons}
\label{Wilson lines}

In this section, we bring in Wilson lines of soft and scalar polarized collinear gluons. We will show that effects of soft gluons and scalar-polarized collinear gluons can be absorbed into these Wilson lines.

We have proved in last section that one can deform the integral path to avoid Glauber region in couplings between soft gluons and collinear particles  in (\ref{propability}). We can thus take the Grammer-Yennie approximation in these couplings. We make such approximation and write the classical effective Lagrangian density in the region $S(x,1/\Lambda_{QCD})$ as:
\begin{equation}
\mathcal{L}_{eff}\equiv\sum_{n^{\mu}}\mathcal{L}_{n}^{(0)}+\mathcal{L}_{s}
\end{equation}
\begin{eqnarray}
\mathcal{L}_{n}^{(0)}&\equiv&i\bar{\psi}_{n}^{(0)}(\widetilde{\not\!\partial}_{n}
-ig\not\!A_{n}^{(0)})\psi_{n}^{(0)}
\nonumber\\
      &&+\frac{1}{2g^{2}}
      tr\left\{([\widetilde{\partial}_{n}^{\phantom{n}\mu}-igA_{n}^{(0)\mu}, \widetilde{\partial}_{n}^{\phantom{n}\nu}-igA_{n}^{(0)\nu}])^{2}\right\}
\end{eqnarray}
\begin{eqnarray}
\mathcal{L}_{s}&\equiv&i\bar{\psi}_{s}(\not\!\partial-ig\not\!A_{s})\psi_{s}
\nonumber\\
&&
+\frac{1}{2g^{2}}tr_{c}\{[\partial^{\mu}-igA_{s}^{\mu},
\partial^{\nu}-igA_{s}^{\nu}]^{2}\}
\end{eqnarray}
where the effective fields $\psi_{n}^{(0)}$ and $A_{n}^{(0)\mu}$ are defined as:
\begin{equation}
\psi_{n}^{(0)}(y_{n})\equiv Y_{n}^{\dag}\psi_{n}(y_{n})
,\quad
A_{n}^{(0)\mu}\equiv Y_{n}^{\dag}A_{n}^{\mu}Y_{n}(y_{n})
\end{equation}
\begin{equation}
\label{soft WL}
Y_{n}(y_{n})=\left\{
   \begin{array}{ll}
     P\exp(ig\int_{-\infty}^{0}\ud s n\cdot A_{s}(y_{n}+sn))
    &\textrm{for $n^{\mu}=n_{1}^{\mu}$ or $n^{\mu}=n_{2}^{\mu}$ }\\
     (P\exp(ig\int_{0}^{\infty}\ud s n\cdot A_{s}(y_{n}+sn)))^{\dag}
     &\textrm{for $n^{\mu}\neq n_{1}^{\mu}$ and $n^{\mu}\neq n_{2}^{\mu}$ }
     \end{array}
     \right.
\end{equation}
, the derivative $\widetilde{\partial}_{n}^{\phantom{\mu}}$ is defined as:
\begin{equation}
\label{partial1}
\widetilde{\partial}_{n}^{\phantom{n}\mu}\psi_{m}^{(0)}\equiv\partial^{\mu}\psi_{m}^{(0)}, \quad
\widetilde{\partial}_{n}^{\phantom{n}\mu}A_{m}^{(0)\nu}\equiv\partial^{\mu}A_{m}^{(0)\nu}
\end{equation}
\begin{equation}
\label{partial2}
\widetilde{\partial}_{n}^{\phantom{n}\mu} A_{s}^{\nu}\equiv\bar{n}^{\mu}n\cdot\partial  A_{s}^{\nu}
\end{equation}
for arbitrary $n^{\mu}$ and $m^{\mu}$.

We extract the large momenta components of collinear fields. That is:
\begin{equation}
\psi_{n}^{(0)}(x_{n})=\sum_{\bar{n}\cdot p}\psi_{n,\bar{n}\cdot p}^{(0)}(x_{n})e^{-i\bar{n}\cdot p n\cdot x_{n}}
\end{equation}
\begin{equation}
A_{n}^{(0)\mu}(x_{n})=\sum_{\bar{n}\cdot p}A_{n,\bar{n}\cdot p}^{(0)\mu}(x_{n})e^{-i\bar{n}\cdot p n\cdot x_{n}}
\end{equation}
Then the large momenta components become labels on the effective fields.(\cite{SCET:2001,SCET:2002,zhou:2013dr,BFPRS:2002})
The classical Lagrangian density in the region $S(x,1/\Lambda_{QCD})$ can then be written as:
\begin{equation}
\label{Leff}
\mathcal{L}_{eff}\equiv\sum_{n^{\mu}}\mathcal{L}_{n}^{(0)}+\mathcal{L}_{s}
\end{equation}
\begin{eqnarray}
\label{Ln}
\mathcal{L}_{n}^{(0)}(y_{n})&\equiv&
i\sum_{\bar{n}\cdot p_{i}}
\bar{\psi}_{n,\bar{n}\cdot p_{1}}^{(0)}
e^{i\bar{n}\cdot p_{1}n\cdot y_{n}}
(\widetilde{\not\!\partial}_{n}
\nonumber\\
&&
-ig\not\!A_{n,\bar{n}\cdot p_{3}}^{(0)}e^{i\bar{n}\cdot p_{3}n\cdot y_{n}})
\psi_{n,\bar{n}\cdot p_{2}}e^{i\bar{n}\cdot p_{3}n\cdot y_{n}}
\nonumber\\
&&+\frac{1}{2g^{2}}\sum_{\bar{n}\cdot p_{i}}
tr\left\{\right.
[
\widetilde{\partial}_{n}^{\phantom{n}\mu}
-igA_{n,\bar{n}\cdot p_{1}}^{(0)\mu}e^{i\bar{n}\cdot p_{1}n\cdot y_{n}},
\nonumber\\
&&
\widetilde{\partial}_{n}^{\phantom{n}\nu}
-igA_{n,\bar{n}\cdot p_{2}}^{(0)\nu}e^{i\bar{n}\cdot p_{2}n\cdot y_{n}}]
\nonumber\\
&&
[\widetilde{\partial}_{n\mu}
-igA_{n,\bar{n}\cdot p_{3}\mu}^{(0)}e^{i\bar{n}\cdot p_{3}n\cdot y_{n}},
\nonumber\\
&&
\left.
\widetilde{\partial}_{n\nu}
-igA_{n,\bar{n}\cdot p_{4}\nu}^{(0)}e^{i\bar{n}\cdot p_{4}n\cdot y_{n}}]
\right\}
\end{eqnarray}
\begin{eqnarray}
\label{Ls}
\mathcal{L}_{s}&\equiv&i\bar{\psi}_{s}(\not\!\partial-ig\not\!A_{s})\psi_{s}
\nonumber\\
&&
+\frac{1}{2g^{2}}tr_{c}\{[\partial^{\mu}-igA_{s}^{\mu},
\partial^{\nu}-igA_{s}^{\nu}]^{2}\}
\end{eqnarray}

We then quantize $\mathcal{L}_{eff}$ by quantizing the effective fields $\psi_{n,\bar{n}\cdot p}^{(0)}$, $A_{n,\bar{n}\cdot p}^{(0)}$, $\psi_{s}$ and $A_{s}$. Hadrons collinear to $n^{\mu}$ move along the direction $n^{\mu}$ with uncertainties of order $1/\Lambda_{QCD}$. Thus boundary conditions in the path integral produced by the initial or final collinear hadrons are constraints on classical configurations of the fields $\psi_{n}^{(0)}(x_{n})$ and $A_{n}^{(0)\mu}(x_{n})$ at $\bar{n}\cdot x_{n}\to\pm\infty $. Boundary conditions produced by final soft hadrons are constraints on classical configurations of the fields $\psi_{s}(x_{s})$ and $A_{s}^{(\mu)}(x_{s})$ at $x_{s}^{0}\to \infty $.

We bring in an effective vertex $\Gamma(x)$ to describe the hard-subprocess in the region $H(x,1/Q)$. We have:
\begin{eqnarray}
\Gamma(x)&=&\Gamma(Y_{n}\psi_{n,\bar{n}\cdot p_{n}}^{(0)}(x),\ldots, Y_{m}A_{m,\bar{m}\cdot p_{m}}^{(0)}Y_{m}^{\dag}(x))
\end{eqnarray}
The residual momenta of the fields $\psi_{n,\bar{n}\cdot p_{n}}^{(0)}$ and $A_{m,\bar{m}\cdot p_{m}}^{(0)}$ only contribute to the momenta conversation of the whole process. We thus set the coordinates of the fields $\psi_{n,\bar{n}\cdot p_{n}}^{(0)}$ and $A_{m,\bar{m}\cdot p_{m}}^{(0)}$ as equal to $x$.

We pause here to discuss the modes $A_{n,\bar{n}\cdot p}^{(0)}$ with $\bar{n}\cdot p=0$.
In usual perturbative calculations, one preform the  integral over the whole momentum space. Thus one can not distinguish the modes $A_{n,\bar{n}\cdot p=0}^{(0)}$ and $A_{s}$ in perturbative calculations.
We absorb contributions of the modes $A_{n,\bar{n}\cdot p=0}^{(0)}$ into those of the modes $A_{s}$. The modes $A_{n,\bar{n}\cdot p=0}^{(0)}$ are not gauge invariant. Thus this should be performed in a given gauge condition, for example, the Feynman gauge.  We make the decomposition:
\begin{eqnarray}
\label{decomp}
A_{n,\bar{n}\cdot p=0}^{(0)\mu}(x_{n})
&=&
n\cdot A_{n,\bar{n}\cdot p=0}^{(0)}(n\cdot x,\bar{n}\cdot x_{n},(\vec{x})_{n\perp})\bar{n}^{\mu}
\nonumber\\
&&
+ A_{n,\bar{n}\cdot p=0}^{(0)\mu}
-n\cdot A_{n,\bar{n}\cdot p=0}^{(0)}(n\cdot x,\bar{n}\cdot x_{n},(\vec{x})_{n\perp})\bar{n}^{\mu}
\end{eqnarray}
where the subscript $n\perp$ denotes that the vector $(\vec{x})_{n\perp}$ fulfil the condition:
\begin{equation}
(\vec{x})_{n\perp}\cdot\vec{n}=0
\end{equation}
Then contribution of the modes $n\cdot A_{n,\bar{n}\cdot p=0}^{(0)}(n\cdot x,\bar{n}\cdot x_{n},(\vec{x})_{n\perp})\bar{n}^{\mu}$ can be absorbed into Wilson lines of the soft gluons.   Contributions of the modes $A_{n,\bar{n}\cdot p=0}^{(0)}(x_{n})
-n\cdot A_{n,\bar{n}\cdot p=0}^{(0)}(n\cdot x,\bar{n}\cdot x_{n},(\vec{x})_{n\perp})\bar{n}^{\mu}$ are power suppressed as the transition probability (\ref{propability}) is free of pinch singular surfaces in Glauber region at leading order in $\Lambda_{QCD}/Q$. We absorb them to collinear modes.   It is equivalent to make the substitution:
\begin{equation}
\label{substitution}
A_{n,\bar{n}\cdot p=0}^{(0)\mu}(x_{n})\to A_{n,\bar{n}\cdot p=0}^{(0)\mu}
-n\cdot A_{n,\bar{n}\cdot p=0}^{(0)}(n\cdot x,\bar{n}\cdot x_{n},(\vec{x})_{n\perp})\bar{n}^{\mu}
\end{equation}
in (\ref{Ln}). This is what we do in Sec.\ref{cancellation}.

We then consider the evolution of the jets collinear to $n^{\mu}$ with the coordinates $\bar{n}\cdot x_{n}$. According to the similar procedure in last section. We see that effects of couplings between particles collinear to $n_{i}^{\mu}$($i=1,2$) occur at the points $x_{n_{i}}$ with $\bar{n_{i}}\cdot x_{n_{i}}>\bar{n_{i}}\cdot x$ cancel out. While effects of couplings between particles collinear to $n^{\mu}$($n^{\mu}\neq n_{1}^{\mu},n^{\mu}\neq n_{2}^{\mu}$) occur at the points $x_{n}$ with $\bar{n}\cdot x_{n}<\bar{n}\cdot x$ cancel out. Thus energy flow from the jets collinear to $n_{i}^{\mu}$($i=1,2$) to the hard vertex and then to the jets collinear to $n^{\mu}$($n^{\mu}\neq n_{1}^{\mu}, n^{\mu}\neq n_{2}^{\mu}$).

We define the momenta of collinear particles that participate in the hard-subprocess as flow from the jets collinear to $n_{i}^{\mu}$($i=1,2$) to the hard vertex or flow from the hard vertex to the jets collinear to $n^{\mu}$($n^{\mu}\neq n_{1}^{\mu}, n^{\mu}\neq n_{2}^{\mu}$). Then we have $\bar{n_{i}}\cdot p_{n_{i}}\geq 0$($i=1,2$) and $\bar{n}\cdot p_{n}\geq 0$($n^{\mu}\neq n_{1}^{\mu}, n^{\mu}\neq n_{2}^{\mu}$), where $p_{n_{i}}$ and $p_{n}$ represent particles collinear to $n_{i}^{\mu}$ and $n^{\mu}$ that participate in the hard-subprocess. To see this, we consider a particle collinear-to-plus with momentum $l$ that participate in the hard sub-process. We repeat the similar calculations as in last section. At the end of this particle that connect to the jet collinear-to-plus, we get the factor:
\begin{equation}
\frac{2i(2\pi)\delta(-l^{+}+\ldots)}{-l^{-}+\ldots+i\epsilon}
(2\pi)^{2}\delta^{(2)}(-\vec{l}_{\perp}+\ldots)
e^{-i\frac{1}{\sqrt{2}}(-l^{+}+\ldots)x^{0}}
\end{equation}
where the ellipsis denotes terms that is independent of $l$.
The denominator of the propagator of $l$ is:
\begin{eqnarray}
l^{2}
&=& 2l^{+}(l^{-}-\frac{|\vec{l}_{\perp}|^{2}}{l^{+}}+i\epsilon l^{+})
\end{eqnarray}
The other end of $l$ is the hard vertex and is independent of $l^{-}$. We see that singular points of $l^{-}$ all locate in the upper half plane if $l^{+}<0$. Thus we have $l^{+}\geq 0$. For particles collinear to other directions, we can repeat the similar calculations and get the conclusion that $\bar{n_{i}}\cdot p_{n_{i}}\geq 0$($i=1,2$) and $\bar{n}\cdot p_{n}\geq 0$($n^{\mu}\neq n_{1}^{\mu}, n^{\mu}\neq n_{2}^{\mu}$).

We now consider the effects of scalar polarized collinear gluons in $\Gamma(x)$.  We denote the momentum of the $j$-th scalar polarized gluons in the hard sub-process that collinear to $n^{\mu}$ as $l_{n}^{j}$. $l_{n}^{j}$ is defined as flow from the jets collinear to $n_{i}^{\mu}$($i=1,2$) to the hard vertex or flow from the hard vertex to the jets collinear to $n^{\mu}$($n^{\mu}\neq n_{1}^{\mu}, n^{\mu}\neq n_{2}^{\mu}$). We work in Feynman gauge in this section thus the other end of $l_{n}^{j}$ should be polarized in the direction $\bar{n}^{\mu}$ at leading order in $\Lambda_{QCD}/Q$. According to our decomposition in (\ref{decomp}) and the substitution in (\ref{substitution}), contribution of these modes are power suppressed if $\bar{n}\cdot l_{n}^{j}=0$.  We thus constrain that $\bar{n}\cdot l_{n}^{j}>0$. We can drop the $i\epsilon$ term in the eikonal propagators. That is to say:
\begin{equation}
\frac{1}{\bar{n}\cdot l_{n}^{j}\pm i\epsilon}=P\left(\frac{1}{\bar{n}\cdot l_{n}^{j}}\right)
\end{equation}
\begin{equation}
\frac{1}{\bar{n}\cdot(l_{n}^{1}+\ldots+l_{n}^{j})\pm i\epsilon}=
P\left(\frac{1}{\bar{n}\cdot(l_{n}^{1}+\ldots+l_{n}^{j})}\right)
\end{equation}
where $P$ denote the principal value. Singular points of the type $\bar{n}\cdot l_{n}^{i}=0$ or $\bar{n}\cdot(l_{n}^{1}+\ldots+l_{n}^{i})=0$ do not give contribution to $\Gamma(x)$ at leading order.  We can then apply the Ward identity to $\Gamma(x)$ to absorb the fields $\bar{n}\cdot A_{n}^{(0)}$ into Wilson lines. We bring in the Wilson lines(\cite{SCET:2001,SCET:2002,BFPRS:2002}):
\begin{equation}
\widetilde{W}_{n,x}^{(0)}=\left\{
   \begin{array}{ll}
     \exp\{g\sum_{\bar{n}\cdot p>0}e^{-i\bar{n}\cdot pn\cdot x}\frac{\bar{n}\cdot A_{\bar{n}\cdot p}^{(0)}(x)}{\bar{n}\cdot P_{n}}\}
    &\textrm{for $n^{\mu}=n_{1}^{\mu}$ or $n^{\mu}=n_{2}^{\mu}$ }\\
     \exp\{g\sum_{\bar{n}\cdot p<0}e^{-i\bar{n}\cdot pn\cdot x}\frac{\bar{n}\cdot A_{\bar{n}\cdot p}^{(0)}(x)}{\bar{n}\cdot P_{n}}\}
     &\textrm{for $n^{\mu}\neq n_{1}^{\mu}$ and $n^{\mu}\neq n_{2}^{\mu}$ }
     \end{array}
     \right.
\end{equation}
where the operator $\bar{n}\cdot P_{n}$ is defined as:
\begin{equation}
\bar{n}\cdot P_{n}\psi_{n,\bar{n}\cdot p}^{(0)}=\bar{n}\cdot p \psi_{n,\bar{n}\cdot p}^{(0)}
,\quad
\bar{n}\cdot P_{n}A_{n,\bar{n}\cdot p}^{(0)\mu}=\bar{n}\cdot p A_{n,\bar{n}\cdot p}^{(0)\mu}
\end{equation}
\begin{equation}
\bar{n}\cdot P_{n}\psi_{m,\bar{n}\cdot p}^{(0)}=\bar{n}\cdot P_{n} \psi_{s}
=
\bar{n}\cdot P_{n}A_{m,\bar{n}\cdot p}^{(0)\mu}=\bar{n}\cdot P_{n} A_{s}^{\mu}
=0(m^{\mu}\neq n^{\mu})
\end{equation}

We notice that contributions of the modes $\bar{n}\cdot A_{\bar{n_{i}}\cdot p}^{(0)}(x)$ with $\bar{n_{i}}\cdot p=0$($i=1,2$) and  $\bar{n}\cdot A_{\bar{n}\cdot p}^{(0)}(x)$($n^{\mu}\neq n_{1}^{\mu},n^{\mu}\neq n_{2}^{\mu}$) with $\bar{n}\cdot p=0$ are power suppressed. We may tentatively relax the constraints $\bar{n_{i}}\cdot p>0$($i=1,2$) and  $\bar{n}\cdot p<0$($n^{\mu}\neq n_{1}^{\mu},n^{\mu}\neq n_{2}^{\mu}$) in the Wilson lines to  that $\bar{n_{i}}\cdot p\geq 0$($i=1,2$) and  $\bar{n}\cdot p\leq 0$($n^{\mu}\neq n_{1}^{\mu},n^{\mu}\neq n_{2}^{\mu}$).  The operators $\bar{n}\cdot P_{n}$ are singular if we include the  modes $\bar{n}\cdot A_{\bar{n}\cdot p}^{(0)}(x)$ with $\bar{n}\cdot p=0$. Thus we need some regularization of the operators $\bar{n}\cdot P_{n}$. It is convenient to regularize these operators so that singular points of the type $\bar{n}\cdot p=0$ do not affect the form of $\Gamma(x)$.

We consider a gluon collinear to plus with momentum $l$ that participate in the hard subprocess. We have:
\begin{equation}
P\left(\frac{1}{l^{+}}\right)=\frac{1}{2}\left(\frac{1}{l^{+}+i\epsilon}+\frac{1}{l^{+}-i\epsilon}\right)
\end{equation}
We repeat the calculations in above paragraphs. At the end of $l$ that connect to the jet collinear-to-plus, we get the factor:
\begin{equation}
\frac{2i(2\pi)\delta(-l^{+}+\ldots)}{-l^{-}+\ldots+i\epsilon}
(2\pi)^{2}\delta^{(2)}(-\vec{l}_{\perp}+\ldots)
e^{-i\frac{1}{\sqrt{2}}(-l^{+}+\ldots)x^{0}}
\end{equation}
where the ellipsis denotes terms that is independent of $l$.
The denominator of the propagator of $l$ is:
\begin{eqnarray}
l^{2}
&=& 2l^{+}(l^{-}-\frac{|\vec{l}_{\perp}|^{2}}{l^{+}}+i\epsilon l^{+})
\end{eqnarray}
The other end of $l$ is the hard vertex and is independent of $l^{-}$. We see that if $l^{+}=-i\epsilon$ then all singular points of $l^{-}$ locate in the upper half plane. In this case, the singular point  $l^{+}=-i\epsilon$ does not affect $\Gamma(x)$. While the singular point $l^{+}=i\epsilon$ does affect $\Gamma(x)$.    Thus we regularize $\bar{n_{1}}\cdot P_{n_{1}}$ in the Wilson lines by making the substitution:
\begin{equation}
\frac{1}{\bar{n_{1}}\cdot P_{n_{1}}}\to \frac{1}{\bar{n_{1}}\cdot P_{n_{1}}+i\epsilon}
\end{equation}
in the Wilson lines. According to the similar analysis, we make the substitution:
\begin{equation}
\frac{1}{\bar{n_{i}}\cdot P_{n_{i}}}\to \frac{1}{\bar{n_{i}}\cdot P_{n_{i}}+i\epsilon}(i=1,2)
\end{equation}
\begin{equation}
\frac{1}{\bar{n}\cdot P_{n}}\to \frac{1}{\bar{n}\cdot P_{n}-i\epsilon}(n^{\mu}\neq n_{1}^{\mu},n^{\mu}\neq n_{2}^{\mu})
\end{equation}
in the Wilson lines.

The modes $\bar{n}\cdot A_{\bar{n_{i}}\cdot p}^{(0)}(x)$ with $\bar{n_{i}}\cdot p<0$($i=1,2$) and  $\bar{n}\cdot A_{\bar{n}\cdot p}^{(0)}(x)$($n^{\mu}\neq n_{1}^{\mu},n^{\mu}\neq n_{2}^{\mu}$) with $\bar{n}\cdot p>0$ do not contribute to the process at leading order according to the analysis in above paragraphs. We can thus drop the constraints $\bar{n_{i}}\cdot p\geq 0$($i=1,2$) and  $\bar{n}\cdot p\leq 0$($n^{\mu}\neq n_{1}^{\mu},n^{\mu}\neq n_{2}^{\mu}$) in the Wilson lines. We can then write the Wilson lines as:
\begin{equation}
\label{collinear WL}
W_{n,x}^{(0)}=\left\{
   \begin{array}{ll}
     P\exp\{ig\int_{-\infty}^{0}\ud s \sum_{\bar{n}\cdot p}e^{-i\bar{n}\cdot p(n\cdot x+s)}\bar{n}\cdot A_{\bar{n}\cdot p}^{(0)}(x)\}
    &\textrm{for $n^{\mu}=n_{1}^{\mu}$ or $n^{\mu}=n_{2}^{\mu}$ }\\
    ( P\exp\{ig\int_{0}^{\infty}\ud s \sum_{\bar{n}\cdot p}e^{-i\bar{n}\cdot p(n\cdot x+s)}\bar{n}\cdot A_{\bar{n}\cdot p}^{(0)}(x)\})^{\dag}
     &\textrm{for $n^{\mu}\neq n_{1}^{\mu}$ and $n^{\mu}\neq n_{2}^{\mu}$ }
     \end{array}
     \right.
\end{equation}

We then bring in  the effective fields:
\begin{eqnarray}
\widehat{\psi}_{n,x}^{(0)}(x_{n})&=&
W_{n,x}^{(0)\dag}\psi_{n}^{(0)}(x_{n})
\nonumber\\
(\widetilde{\partial}_{n}^{\phantom{n}\mu}-ig\widehat{A}_{n,x}^{(0)\mu})(x_{n})&=&
W_{n,x}^{(0)\dag}
(\widetilde{\partial}_{n}^{\phantom{n}\mu}-igA_{n}^{(0)\mu})
W_{n,x}^{(0)}
(x_{n})
\end{eqnarray}
$\Gamma(x)$ can then be written as:
\begin{eqnarray}
\label{hard vertex}
\Gamma(x)
&=&
\sum_{n,\bar{n}\cdot p,\ldots,m,\bar{m}\cdot p^{\prime}}
\Gamma^{\mu}(Y_{n}(\widehat{\psi}_{n,x}^{(0)})_{\bar{n}\cdot p},\ldots,
\nonumber\\
&&
Y_{m}(\widetilde{\partial}_{m}^{\phantom{m}m\perp}-ig\widehat{A}_{n,x}^{(0)m\perp})_{\bar{m}\cdot p^{\prime}}
Y_{m}^{\dag})(x)
\end{eqnarray}
where the short distance coefficients in $\Gamma$ are functions of the large momenta components $\bar{n}\cdot p_{n}$, $\vec{m\perp}$ denotes the vectors that fulfill the condition $\vec{m\perp}\cdot \vec{m}=0$.
Physical fields in $\Gamma^{\mu}$ should connect to different jets at leading order.(\cite{zhou:2013dr,S:1978}). Thus $\Gamma^{\mu}$ is multi-linear with its variables.

Up to this step, we are dealing with the Wilson lines structures in Hadron-Hadron collision. Our conclusion is that effects of soft gluons and scalar polarized collinear gluons can be absorbed into Wilson lines of these gluons. The remain question is whether the non-trivial dependence of cross section on transverse momenta of exchanged gluons in (\cite{Roggers:2013}) really break the factorization. We will consider this in next section.
\\

\section{Nontrivial Transverse Momenta Dependence of Hadron-Hadron Collisions }
\label{tr dependence}

In this section we inspect the nontrivial transverse momenta dependence of hadron-hadron collisions  shown in \cite{Roggers:2013}. Our conclusion is that such non-trivial transverse momenta  dependence can be described by the effective hard vertex (\ref{hard vertex}).

\begin{figure*}
\centering
\includegraphics[scale=0.5]{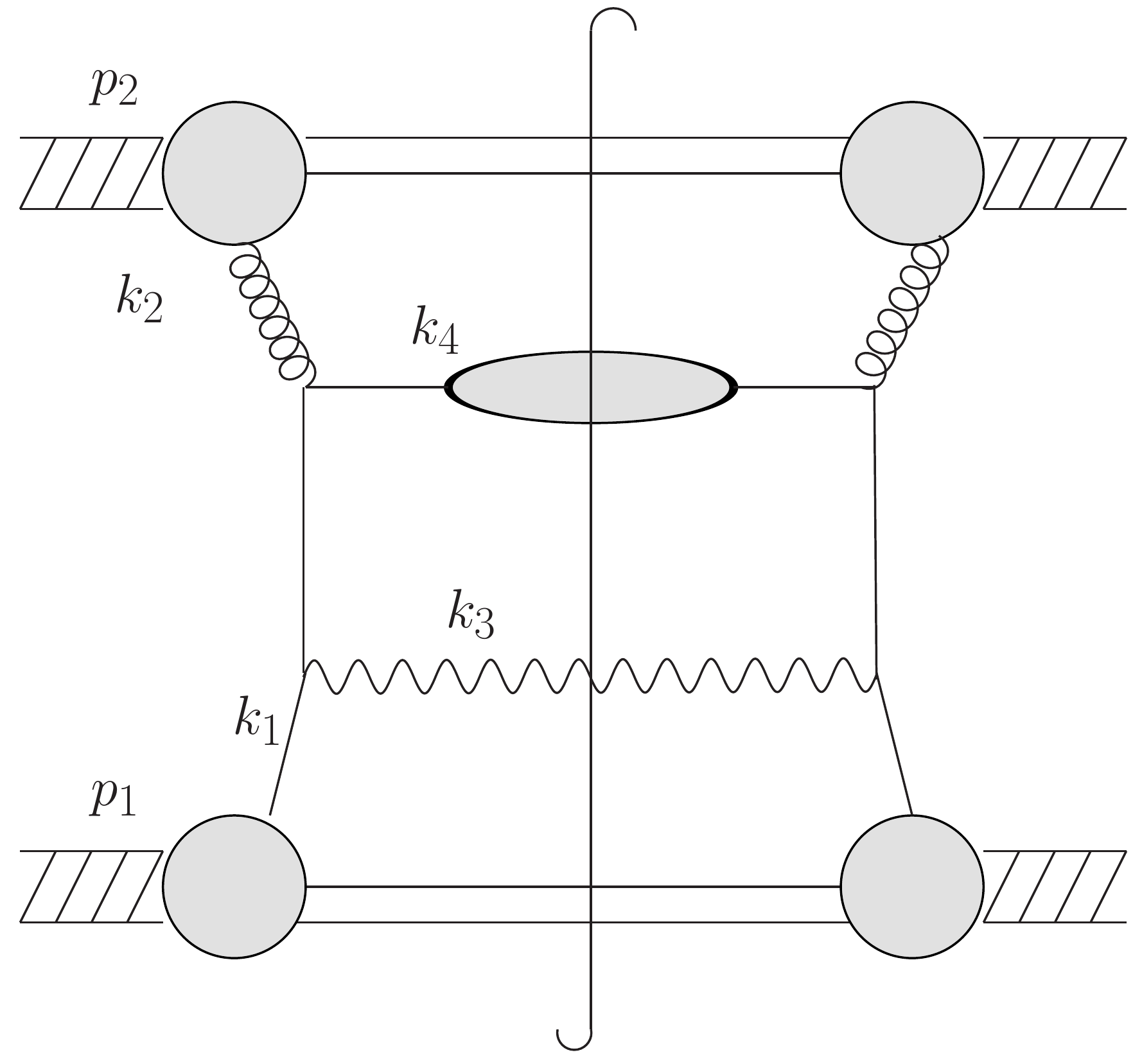}
\caption{An example of the inclusive production of hadron-photon pair in hadron-hadron collision at tree level}
\label{photon}
\end{figure*}

To compare our results with those in (\cite{Roggers:2013}), we consider the inclusive production of a back-to-back hadron-photon pair in hadron-hadron collision(see Fig.\ref{photon} ):
\begin{equation}
p_{1}+p_{2}\to \gamma(k_{3},\lambda_{3})+H_{4}(k_{3}, S_{4})+X
\end{equation}
where $p_{1}$ and $p_{2}$ represent initial hadrons, $\gamma$ represents the detected final photon, $H_{4}$ represents the detected final hadron,  $X$ represents any other particles, $k_{3}$ and  $\lambda_{3}$ denote the momentum and helicity of the detected photon, $k_{4}$ and  $S_{4}$ denote the momentum and spin of the detected hadron.

$k_{3}$ and $k_{4}$ are constrainers to be not nearly collinear to $p_{1}$ or $p_{2}$. The hard energy scale of this process is set by the quantity:
\begin{equation}
Q\equiv\sqrt{(k_{3}+k_{4})^{2}}
\end{equation}

As in (\cite{Roggers:2013}), we consider the $gq\to\gamma q$ channel in this paper. An example of such channel at tree level is shown in Fig.\ref{photon}.

We work in the center of mass frame of initial hadrons. The momenta $p_{1}$ and $p_{2}$ are approximately light-like and denoted as:
\begin{equation}
p_{1}\simeq (p_{1}^{+},0,\vec{0})
,\quad
p_{2}\simeq (0,p_{2}^{-},\vec{0})
\end{equation}
We bring in the notations:
\begin{equation}
n_{1}^{\mu}\equiv\frac{1}{\sqrt{2}}(1,0,0,1)=\frac{1}{\sqrt{2}}(1,0,\vec{0})
\end{equation}
\begin{equation}
n_{2}^{\mu}\equiv\frac{1}{\sqrt{2}}(1,0,0,-1)=\frac{1}{\sqrt{2}}(0,1,\vec{0})
\end{equation}
\begin{equation}
\bar{n_{i}}^{\mu}\equiv\sqrt{2}(1,0,0,0)-n_{i}^{\mu}(i=1,2)
\end{equation}
then we have:
\begin{equation}
p_{1}^{\mu}\simeq \bar{n}_{1}\cdot p_{1} n_{1}^{\mu}
,\quad
p_{2}^{\mu}\simeq \bar{n}_{2}\cdot p_{2} n_{2}^{\mu}
\end{equation}
The incoming partons are approximately collinear to their parent hadrons, thus $k_{1}$ and $k_{2}$ scale as:
\begin{equation}
(k_{1}^{+},k_{1}^{-},|\vec{k_{1}}_{\perp}|)\sim (Q,\Lambda_{QCD}^{2}/Q,\Lambda_{QCD})
,\quad
(k_{2}^{+},k_{2}^{-},|\vec{k_{2}}_{\perp}|)\sim (\Lambda_{QCD}^{2}/Q,Q,\Lambda_{QCD})
\end{equation}
$k_{3}$ and $k_{4}$ are also collinear at the limit $Q\to\infty$.  they scale as:
 \begin{equation}
(k_{3}^{+},k_{3}^{-},|\vec{k_{3}}_{\perp}|)\sim (Q,Q,Q)
,\quad
(k_{4}^{+},k_{4}^{-},|\vec{k_{4}}_{\perp}|)\sim (Q,Q,Q)
\end{equation}
We bring in the notation:
\begin{equation}
n_{3}^{\mu}\equiv\frac{1}{\sqrt{2}}(1,\vec{n_{3}})
\end{equation}
\begin{equation}
n_{4}^{\mu}\equiv\frac{1}{\sqrt{2}}(1,\vec{n_{4}})
\end{equation}
where
\begin{equation}
|\vec{n_{3}}|=|\vec{n_{4}}|=1
\end{equation}
\begin{equation}
n_{3}\cdot n_{4}=1+O(|\vec{q}_{\perp}|/Q)
\end{equation}
We have:
\begin{equation}
k_{3}^{\mu}\simeq \bar{n}_{3}\cdot k_{3} n_{3}^{\mu}
,\quad
k_{4}^{\mu}\simeq \bar{n}_{4}\cdot k_{2} n_{4}^{\mu}
\end{equation}
where
\begin{equation}
\bar{n_{i}}^{\mu}\equiv\sqrt{2}(1,0,0,0)-n_{i}^{\mu}
\end{equation}

\subsection{Calculations of Hard Sub-process Involving One Gluon }

\begin{figure*}
\centering
\begin{tabular}{c@{\hspace*{10mm}}c}
\includegraphics[scale=0.3]{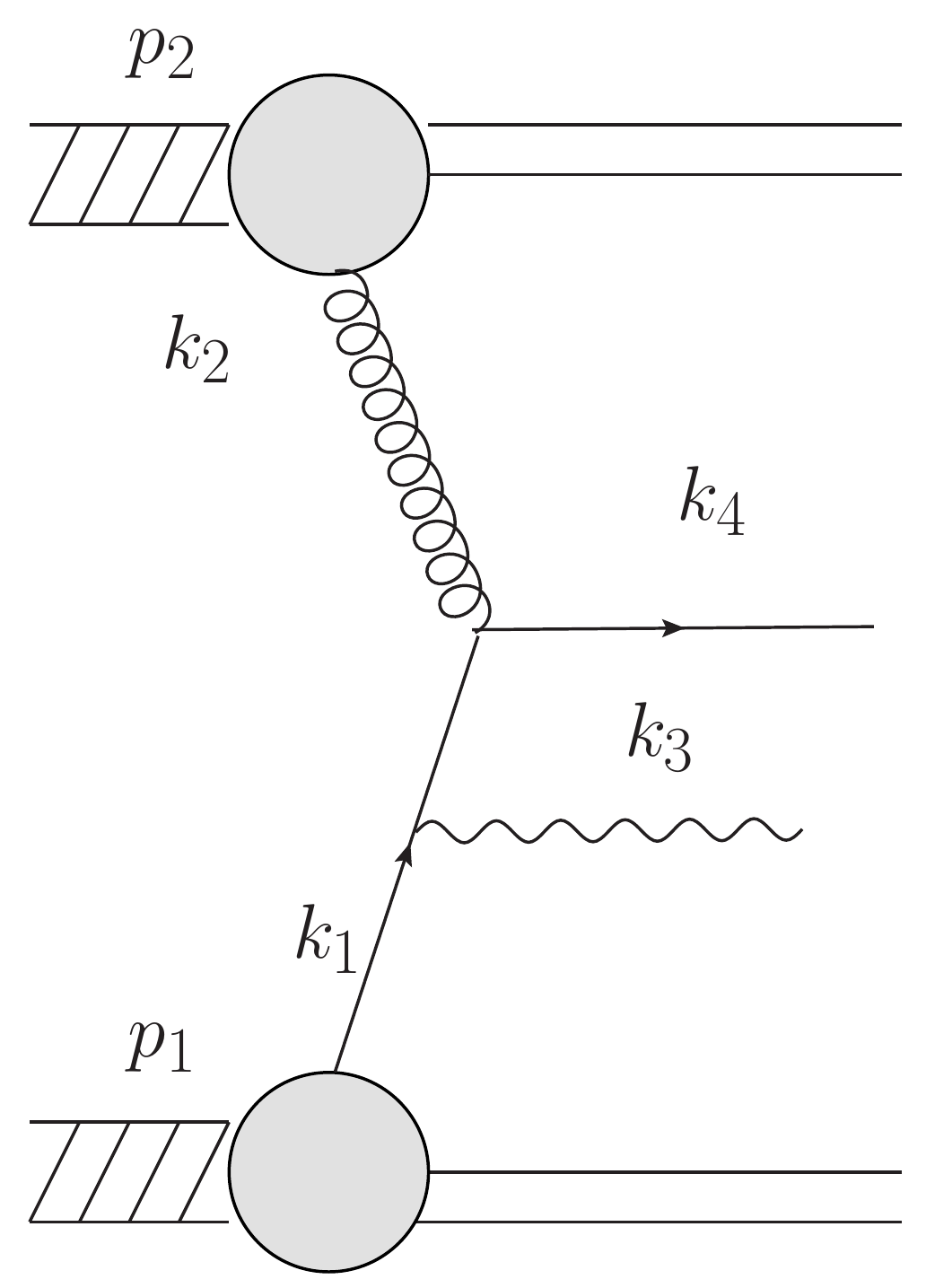}
&
\includegraphics[scale=0.3]{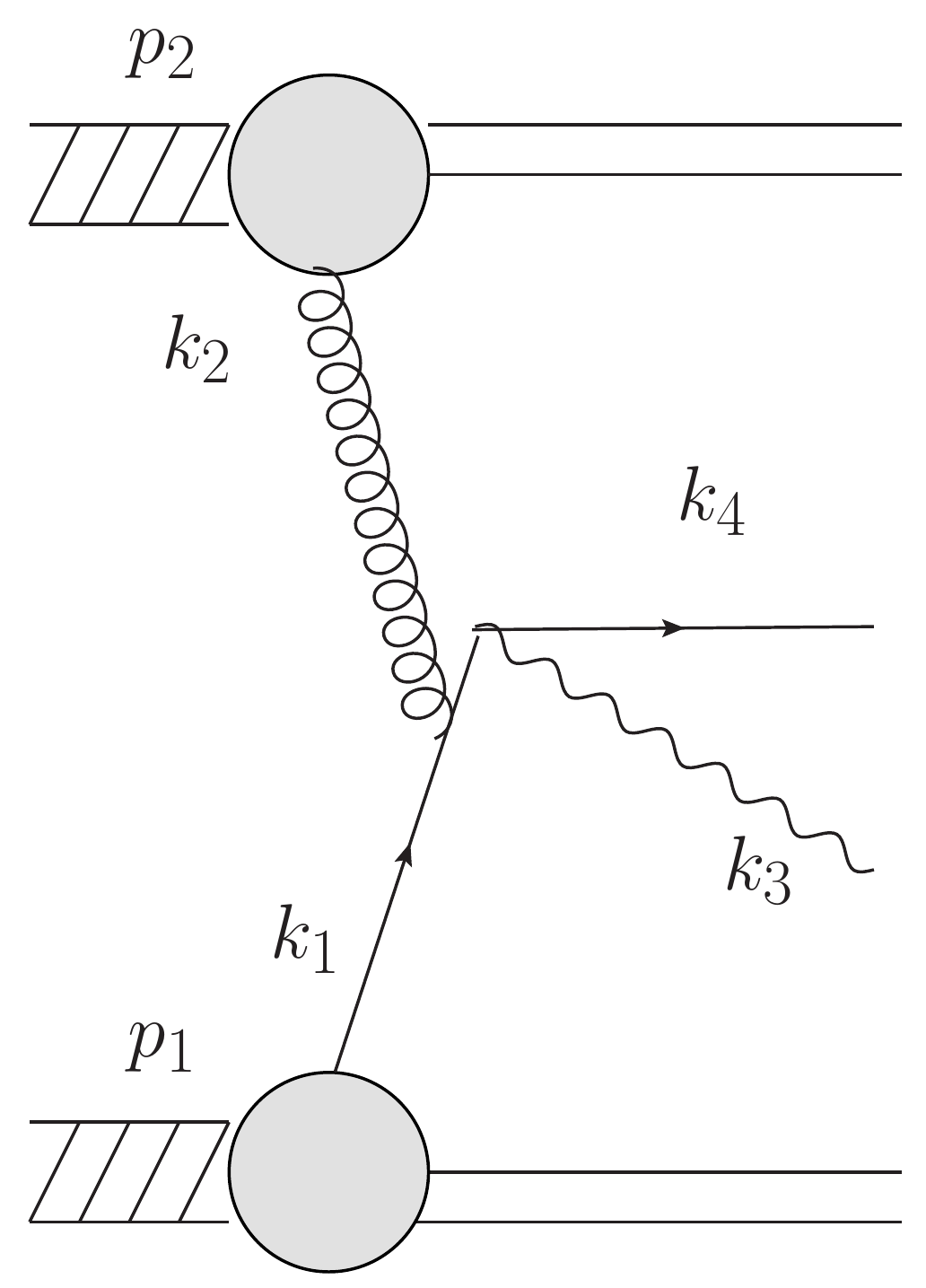}
\\
(a)&(b)
\end{tabular}
\caption{Inclusive production of a back-to-back hadron-photon pair with one initial gluon at tree level}
\label{oneg}
\end{figure*}

We start from the case that there is only one gluon participating in the hard sub-process as shown in
Fig.(\ref{oneg}).

The on-shell scattering amplitude at parton level is:
\begin{eqnarray}
iM_{on-shell}&\equiv&\bar{u}_{S_{4}}(k_{4})\not\!\epsilon^{a}(k_{2})(igt^{a})\frac{i(\not\!k_{4}-\not\! k_{2})}{(k_{4}-k_{2})^{2}+i\epsilon}\not\!\epsilon_{\lambda_{3}}^{*}(k_{3})(iQ_{q}e)u(k_{1})
\nonumber\\
&&+\bar{u}_{S_{4}}(k_{4})\not\!\epsilon_{\lambda_{3}}^{*}(k_{3})(iQ_{q}e)\frac{i(\not\!k_{1}+\not\! k_{2})}{(k_{1}+k_{2})^{2}+i\epsilon}\not\!\epsilon^{a}(k_{2})(igt^{a})u(k_{1})
\end{eqnarray}
where $Q_{q}$ is the electric-charge of the quark $k_{2}$. At leading order in $\Lambda_{QCD}/Q$, such scattering amplitude can be reproduced by the effective operator:
\begin{eqnarray}
J^{\mu}B_{\mu}(x)
&\equiv&\sum_{n_{1}^{\mu},\bar{n_{1}}\cdot k_{1}}
\sum_{n_{2}^{\mu},\bar{n_{2}}\cdot k_{2}}
\sum_{n_{4}^{\mu},\bar{n_{4}}\cdot k_{4}}
\bar{\psi}_{n_{4},\bar{n_{4}}\cdot k_{4},x}^{(0)}(-\widetilde{\not\!\partial}_{n_{2}}^{\phantom{n_{2}}n_{2}\perp}+ig\not\! A_{n_{2},\bar{n_{2}}\cdot k_{2},x}^{(0)n_{2}\perp})
\nonumber\\
&&
\frac{i(\bar{n_{4}}\cdot k_{4}\not\!n_{4}-\bar{n_{2}}\cdot k_{2}\not\!n_{2})}{-2\bar{n_{2}}\cdot k_{2}\bar{n_{4}}\cdot k_{4} n_{2}\cdot n_{4}+i\epsilon}
(Q_{q}e\not\! B)\psi_{n_{1},\bar{n_{1}}\cdot k_{1},x}^{(0)}(x)
\nonumber\\
&&+\sum_{n_{1}^{\mu},\bar{n_{1}}\cdot k_{1}}
\sum_{n_{2}^{\mu},\bar{n_{2}}\cdot k_{2}}
\sum_{n_{4}^{\mu},\bar{n_{4}}\cdot k_{4}}
\bar{\psi}_{n_{4},\bar{n_{4}}\cdot k_{4},x}^{(0)}(Q_{q}e\not\! B)
\nonumber\\
&&
\frac{i(\bar{n_{1}}\cdot k_{1}\not\!n_{1}+\bar{n_{2}}\cdot k_{2}\not\!n_{2})}{2\bar{n_{1}}\cdot k_{1}\bar{n_{2}}\cdot k_{2}n_{1}\cdot n_{2}+i\epsilon}
(-\widetilde{\not\!\partial}_{n_{2}}^{\phantom{n_{2}}n_{2}\perp}+ig\not\! A_{n_{2},\bar{n_{2}}\cdot k_{2},x}^{(0)n_{2}\perp})\psi_{n_{1},\bar{n_{1}}\cdot k_{1},x}^{(0)}(x)
\end{eqnarray}
where $B^{\mu}$ denotes the photon field, $\vec{n\perp}$ denotes the vectors that fulfill the condition $\vec{n\perp}\cdot \vec{n}=0$, the derivative $\widetilde{\partial}_{n}^{\phantom{n}\mu}$ is defined in (\ref{partial1}) and (\ref{partial2}).

According to the form of $\Gamma(x)$ in (\ref{hard vertex}), the effective operators that contribute to the process considered here is:
\begin{eqnarray}
\label{effop}
\Gamma^{\mu}B_{\mu}(x)
&\equiv&\sum_{n_{1}^{\mu},\bar{n_{1}}\cdot k_{1}}
\sum_{n_{2}^{\mu},\bar{n_{2}}\cdot k_{2}}
\sum_{n_{4}^{\mu},\bar{n_{4}}\cdot k_{4}}
\nonumber\\
&&
\bar{\widehat{\psi}}_{n_{4},\bar{n_{4}}\cdot k_{4},x}^{(0)}Y_{n_{4}}^{\dag}
Y_{n_{2}}
(-\widetilde{\not\!\partial}_{n_{2}}^{\phantom{n_{2}}n_{2}\perp}+ig\not\! \widehat{A}_{n_{2},\bar{n_{2}}\cdot k_{2},x}^{(0)n_{2}\perp})
Y_{n_{2}^{\dag}}
\nonumber\\
&&
\frac{i(\bar{n_{4}}\cdot k_{4}\not\!n_{4}-\bar{n_{2}}\cdot k_{2}\not\!n_{2})}{-2\bar{n_{2}}\cdot k_{2}\bar{n_{4}}\cdot k_{4} n_{2}\cdot n_{4}+i\epsilon}
(Q_{q}e\not\! B)Y_{n_{1}}\widehat{\psi}_{n_{1},\bar{n_{1}}\cdot k_{1},x}^{(0)}(x)
\nonumber\\
&&+\sum_{n_{1}^{\mu},\bar{n_{1}}\cdot k_{1}}
\sum_{n_{2}^{\mu},\bar{n_{2}}\cdot k_{2}}
\sum_{n_{4}^{\mu},\bar{n_{4}}\cdot k_{4}}
\bar{\widehat{\psi}}_{n_{4},\bar{n_{4}}\cdot k_{4},x}^{(0)}Y_{n_{4}}^{\dag}
(Q_{q}e\not\! B)
\nonumber\\
&&
\frac{i(\bar{n_{1}}\cdot k_{1}\not\!n_{1}+\bar{n_{2}}\cdot k_{2}\not\!n_{2})}{2\bar{n_{1}}\cdot k_{1}\bar{n_{2}}\cdot k_{2}n_{1}\cdot n_{2}+i\epsilon}
\nonumber\\
&&
Y_{n_{2}}
(-\widetilde{\not\!\partial}_{n_{2}}^{\phantom{n_{2}}n_{2}\perp}+ig\not\! \widehat{A}_{n_{2},\bar{n_{2}}\cdot k_{2},x}^{(0)n_{2}\perp})
Y_{n_{2}^{\dag}}
Y_{n_{1}}\widehat{\psi}_{n_{1},\bar{n_{1}}\cdot k_{1},x}^{(0)}(x)
\end{eqnarray}

We then consider the diagrams in Fig.\ref{oneg}. The gluon $k_{2}$ is collinear gluon, thus we can drop the Wilson lines $Y_{n_{i}}$ in (\ref{effop}). we have:
\begin{eqnarray}
\Gamma^{\mu}B_{\mu}(x)|_{\text{one gluon}}
&\equiv&\sum_{n_{1}^{\mu},\bar{n_{1}}\cdot k_{1}}
\sum_{n_{2}^{\mu},\bar{n_{2}}\cdot k_{2}}
\sum_{n_{4}^{\mu},\bar{n_{4}}\cdot k_{4}}
\nonumber\\
&&
\bar{\psi}_{n_{4},\bar{n_{4}}\cdot k_{4},x}^{(0)}
(-\widetilde{\not\!\partial}_{n_{2}}^{\phantom{n_{2}}n_{2}\perp}+ig\not\! A_{n_{2},\bar{n_{2}}\cdot k_{2},x}^{(0)n_{2}\perp}
-ig\frac{(\widetilde{\not\!\partial}_{n_{2}}^{\phantom{n_{2}}n_{2}\perp}\bar{n_{2}}\cdot A_{n_{2},\bar{n_{2}}\cdot k_{2},x}^{(0)})}{(-i)\bar{n_{2}}\cdot k_{2}})
\nonumber\\
&&
\frac{i(\bar{n_{4}}\cdot k_{4}\not\!n_{4}-\bar{n_{2}}\cdot k_{2}\not\!n_{2})}{-2\bar{n_{2}}\cdot k_{2}\bar{n_{4}}\cdot k_{4} n_{2}\cdot n_{4}+i\epsilon}
(Q_{q}e\not\! B)\psi_{n_{1},\bar{n_{1}}\cdot k_{1},x}^{(0)}(x)
\nonumber\\
&&+\sum_{n_{1}^{\mu},\bar{n_{1}}\cdot k_{1}}
\sum_{n_{2}^{\mu},\bar{n_{2}}\cdot k_{2}}
\sum_{n_{4}^{\mu},\bar{n_{4}}\cdot k_{4}}
\bar{\psi}_{n_{4},\bar{n_{4}}\cdot k_{4},x}^{(0)}
(Q_{q}e\not\! B)
\nonumber\\
&&
\frac{i(\bar{n_{1}}\cdot k_{1}\not\!n_{1}+\bar{n_{2}}\cdot k_{2}\not\!n_{2})}{2\bar{n_{1}}\cdot k_{1}\bar{n_{2}}\cdot k_{2}n_{1}\cdot n_{2}+i\epsilon}
\nonumber\\
&&
(-\widetilde{\not\!\partial}_{n_{2}}^{\phantom{n_{2}}n_{2}\perp}+ig\not\! A_{n_{2},\bar{n_{2}}\cdot k_{2},x}^{(0)n_{2}\perp}
-ig\frac{(\widetilde{\not\!\partial}_{n_{2}}^{\phantom{n_{2}}n_{2}\perp}\bar{n_{2}}\cdot A_{n_{2},\bar{n_{2}}\cdot k_{2},x}^{(0)})}{(-i)\bar{n_{2}}\cdot k_{2}})
\psi_{n_{1},\bar{n_{1}}\cdot k_{1},x}^{(0)}(x)
\end{eqnarray}
The $\widetilde{\not\!\partial}_{n_{2}}^{\phantom{n_{2}}n_{2}\perp}$ term does not contribute to scattering amplitude according to momenta conversation.  We make the substitution:
\begin{equation}
A_{n_{2},\bar{n_{2}}\cdot k_{2},x}^{(0)n_{2}\perp\mu}
-\frac{(i\partial^{n_{2}\perp\mu}\bar{n_{2}}\cdot A_{n_{2},\bar{n_{2}}\cdot k_{2},x}^{(0)})}{\bar{n_{2}}\cdot k_{2}})
\to A_{n_{2},\bar{n_{2}}\cdot k_{2},x}^{(0)\mu}
-\frac{(i\partial^{n_{2}\perp\mu}\bar{n_{2}}\cdot A_{n_{2},\bar{n_{2}}\cdot k_{2},x}^{(0)})}{\bar{n_{2}}\cdot k_{2}})
\end{equation}
which do not change the amplitude at leading order according to Ward identity. We bring in the operator:
\begin{eqnarray}
\widetilde{\Gamma}^{\mu}B_{\mu}(x)|_{\text{one gluon}}
&\equiv&\sum_{n_{1}^{\mu},\bar{n_{1}}\cdot k_{1}}
\sum_{n_{2}^{\mu},\bar{n_{2}}\cdot k_{2}}
\sum_{n_{4}^{\mu},\bar{n_{4}}\cdot k_{4}}
\nonumber\\
&&
\bar{\psi}_{n_{4},\bar{n_{4}}\cdot k_{4},x}^{(0)}(ig)
(\not\! A_{n_{2},\bar{n_{2}}\cdot k_{2},x}^{(0)n_{2}}
-\frac{(i\widetilde{\not\!\partial}_{n_{2}}^{\phantom{n_{2}}n_{2}\perp}\bar{n_{2}}\cdot A_{n_{2},\bar{n_{2}}\cdot k_{2},x}^{(0)})}{\bar{n_{2}}\cdot k_{2}})
\nonumber\\
&&
\frac{i(\bar{n_{4}}\cdot k_{4}\not\!n_{4}-\bar{n_{2}}\cdot k_{2}\not\!n_{2})}{-2\bar{n_{2}}\cdot k_{2}\bar{n_{4}}\cdot k_{4} n_{2}\cdot n_{4}+i\epsilon}
(Q_{q}e\not\! B)\psi_{n_{1},\bar{n_{1}}\cdot k_{1},x}^{(0)}(x)
\nonumber\\
&&+\sum_{n_{1}^{\mu},\bar{n_{1}}\cdot k_{1}}
\sum_{n_{2}^{\mu},\bar{n_{2}}\cdot k_{2}}
\sum_{n_{4}^{\mu},\bar{n_{4}}\cdot k_{4}}
\bar{\psi}_{n_{4},\bar{n_{4}}\cdot k_{4},x}^{(0)}
(Q_{q}e\not\! B)
\nonumber\\
&&
\frac{i(\bar{n_{1}}\cdot k_{1}\not\!n_{1}+\bar{n_{2}}\cdot k_{2}\not\!n_{2})}{2\bar{n_{1}}\cdot k_{1}\bar{n_{2}}\cdot k_{2}n_{1}\cdot n_{2}+i\epsilon}
\nonumber\\
&&
(ig)(\not\! A_{n_{2},\bar{n_{2}}\cdot k_{2},x}^{(0)n_{2}}
-\frac{(i\widetilde{\not\!\partial}_{n_{2}}^{\phantom{n_{2}}n_{2}\perp}\bar{n_{2}}\cdot A_{n_{2},\bar{n_{2}}\cdot k_{2},x}^{(0)})}{\bar{n_{2}}\cdot k_{2}})
\psi_{n_{1},\bar{n_{1}}\cdot k_{1},x}^{(0)}(x)
\end{eqnarray}
The scattering amplitude of the process can then be written as:
\begin{equation}
iM\equiv\big<k_{3},\lambda_{3};k_{4},S_{4}|T(\widetilde{\Gamma}^{\mu}B_{\mu})|_{\text{one  gluon}}|p_{1}p_{2}\big>
\end{equation}
where $T$ denotes the time-ordering operator. This is in accordance with the corresponding result in \cite{Roggers:2013}.
\\

\subsection{Calculations of Hard Sub-process Involving Two Gluons }

We now consider the case that there are two gluons connect to the hadron $p_{2}$. There is the case that the additional gluon do not connect to hard sub-diagram. That is to say, the additional gluon is exchanged between spectators. This gluon should be soft at leading order in $\Lambda_{QCD}/Q$. We perform the summation over all possible final cuts of spectators and absorb soft gluons in to the Wilson lines of soft gluons in Sec.\ref{Wilson lines}. Thus we do not show the calculations of such case. This is enough to compare with the results in \cite{Roggers:2013}.

The imaginary parts of eikonal propagators are not concerned in \cite{Roggers:2013}, as they do not violate maximally generalized factorization TMD-factorization in the one-extra-gluon example. To simplify the calculations, we will also drop such terms in this paper. We do not distinguish collinear or soft gluons in these calculations as contributions of the singular points $\bar{n_{i}}\cdot p_{n_{i}}=0$ are not concerned in these calculations, where $p_{n_{i}}$ denotes the momentum of a internal line that is collinear to $n_{i}^{\mu}$. We can thus make the substitution:
\begin{eqnarray}
Y_{n}\widehat{\psi}_{n,x}^{(0)}&\to& \widehat{\psi}_{n,x}
\nonumber\\
Y_{n}(\widetilde{\partial}_{n}^{\phantom{n}\mu}-ig\widehat{A}_{n,x}^{(0)\mu})
Y_{n}^{\dag}
&\to&
(\widetilde{\partial}_{n}^{\phantom{n}\mu}-ig\widehat{A}_{n,x}^{\mu})
\end{eqnarray}
in (\ref{effop}), where
\begin{eqnarray}
\widehat{\psi}_{n,x}(x_{n})&=&W_{n,x}^{\dag}(x_{n})\psi_{n}(x_{n})
\nonumber\\
(\widetilde{\partial}_{n}^{\phantom{n}\mu}-ig\widehat{A}_{n,x}^{\mu})&=&
W_{n,x}^{\dag}
(\widetilde{\partial}_{n}^{\phantom{n}\mu}-igA_{n}^{\mu})
W_{n,x}
\end{eqnarray}
\begin{equation}
W_{n,x}=\left\{
   \begin{array}{ll}
     P\exp\{ig\int_{-\infty}^{0}\ud s \sum_{\bar{n}\cdot p}e^{-i\bar{n}\cdot p(n\cdot x+s)}\bar{n}\cdot A_{\bar{n}\cdot p}(x)\}
    &\textrm{for $n^{\mu}=n_{1}^{\mu}$ or $n^{\mu}=n_{2}^{\mu}$ }\\
    ( P\exp\{ig\int_{0}^{\infty}\ud s \sum_{\bar{n}\cdot p}e^{-i\bar{n}\cdot p(n\cdot x+s)}\bar{n}\cdot A_{\bar{n}\cdot p}(x)\})^{\dag}
     &\textrm{for $n^{\mu}\neq n_{1}^{\mu}$ and $n^{\mu}\neq n_{2}^{\mu}$ }
     \end{array}
     \right.
\end{equation}

We now consider the effective operator $\Gamma^{\mu}B_{\mu}$ in (\ref{effop}) at the two-gluon level(both gluons connect to $p_{2}$). We have:
\begin{eqnarray}
&&\Gamma^{\mu}B_{\mu}|_{\text{two gluons}}(k_{1},k_{2},k_{3},k_{4})
\nonumber\\
&\equiv&-\frac{f^{abc}}{2}g^{2}(Q_{q}e) B^{\mu}(k_{3}))\sum_{n_{1}^{\nu}}
\sum_{n_{2}^{\nu}}
\sum_{n_{4}^{\nu}}
\bar{\psi}_{n_{4}}(k_{4})
\nonumber\\
&&
\left\{-\int\frac{\ud^{4}l_{2}}{(2\pi)^{4}}l_{2}^{n_{2}\perp\nu}
P\left(\frac{\bar{n_{2}}\cdot A_{n}^{a}(k_{2}-l_{2})}{\bar{n_{2}}\cdot (k_{2}-l_{2})}\right)
P\left(\frac{\bar{n_{2}}\cdot A_{n}^{b}(l_{2})}{\bar{n_{2}}\cdot l_{2}}\right)\right.
\nonumber\\
&&
+\int\frac{\ud^{4}l_{2}}{(2\pi)^{4}}k_{2}^{n_{2}\perp\nu}
P\left(\frac{\bar{n_{2}}\cdot A_{n}^{a}(k_{2}-l_{2})}{\bar{n_{2}}\cdot k_{2}}\right)
P\left(\frac{\bar{n_{2}}\cdot A_{n}^{b}(l_{2})}{\bar{n_{2}}\cdot l_{2}}\right)
\nonumber\\
&&
\left.
-2\int\frac{\ud^{4}l_{2}}{(2\pi)^{4}}A_{n_{2}}^{n_{2}\perp\nu a}(k_{2}-l_{2})
P\left(\frac{\bar{n_{2}}\cdot A_{n}^{b}(l_{2})}{\bar{n_{2}}\cdot l_{2}}\right)
\right\}
\nonumber\\
&&
\left\{
\frac{\gamma_{\nu}(\bar{n_{4}}\cdot k_{4}\not\!n_{4}
-\bar{n_{2}}\cdot k_{2}\not\!n_{2})\gamma_{\mu}t^{c}}
{-2\bar{n_{2}}\cdot k_{2}\bar{n_{4}}\cdot k_{4} n_{2}\cdot n_{4}+i\epsilon}\right.
\nonumber\\
&&
\left.
+
\frac{\gamma_{\mu}(\bar{n_{1}}\cdot k_{1}\not\!n_{1}
+\bar{n_{2}}\cdot k_{2}\not\!n_{2})\gamma_{\nu}t^{c}}
{2\bar{n_{1}}\cdot k_{1}\bar{n_{2}}\cdot k_{2}n_{1}\cdot n_{2}+i\epsilon}
\right\}\psi_{n_{1}} (k_{1})
\end{eqnarray}
where we have turn in to momenta space to simplify the expression, $P$ denotes the principal value. The $l_{2}^{n_{2}\perp\nu}$ term is indeed the same as that in the formula (134) of \cite{Roggers:2013}. One should not be confused with the global coefficient $1/2$ in our formula. We do not contract fields in the operator $\Gamma^{\mu}B_{\mu}$ with initial or final states at this step. There are two possible manners to contract the two gluon fields in $\Gamma^{\mu}B_{\mu}$ and the two initial gluons. We have dropped terms with the color factor $\frac{1}{2}\{t^{a},t^{b}\}$ in above formula as they do not contribute to the amplitude of the process considered here.
The scattering amplitude of the process can then be written as:
\begin{equation}
iM\equiv\big<k_{3},\lambda_{3};k_{4},S_{4}|T(\widetilde{\Gamma}^{\mu}B_{\mu})|_{\text{two  gluons}}|p_{1}p_{2}\big>
\end{equation}
This is in accordance with the corresponding result in \cite{Roggers:2013}

We see that the non-trivial transverse momenta dependence of cross section, which is considered as the symbol of break down of factorization, can be described by the effective operator $\widetilde{\Gamma}^{\mu}$ in (\ref{effop}) or more generally, by the effective operator $\Gamma(x)$  in (\ref{hard vertex}). We will work in the frame of effective theory and finish the proof of factorization theorem in next section.
\\

\section{Factorization.}
\label{factorize}

In this section, we finish the proof of QCD factorization for processes considered in this paper.

We start from the transition probability:
\begin{eqnarray}
\label{trprob}
|M|^{2}&\equiv&\lim_{T\to\infty}\sum_{X}\int\ud^{4}x_{2}\int\ud^{4}x_{1}
\nonumber\\
&&
_{-T}\big<p_{1}p_{2}|
e^{iH_{eff}(x_{2}^{0}+T)}
\Gamma^{\dag}(\vec{x}_{2})
e^{iH_{eff}(T-x_{2}^{0})}
|H_{3}H_{4}X\big>_{T}
\nonumber\\
&&
_{T}\big<H_{3}H_{4}X|
e^{-iH_{eff}(T-x_{1}^{0})}
\Gamma(\vec{x}_{1})
e^{-iH_{eff}(x_{1}^{0}+T)}
|p_{1}p_{2}\big>_{-T}
\nonumber\\
&=&
\lim_{T\to\infty}\sum_{X}\int\frac{\ud^{4}k}{(2\pi)^{4}}\int\ud^{4}xe^{-ik\cdot x}
\nonumber\\
&&
_{-T}\big<p_{1}p_{2}|
e^{iH_{eff}(x^{0}+T)}
\Gamma^{\dag}(\vec{x})
e^{iH_{eff}(T-x^{0})}
|H_{3}H_{4}X\big>_{T}
\nonumber\\
&&
_{T}\big<H_{3}H_{4}X|
e^{-iH_{eff}T}
\Gamma(\vec{0})
e^{-iH_{eff}T}
|p_{1}p_{2}\big>_{-T}
\end{eqnarray}
where $\Gamma(x)$ take the form shown in (\ref{hard vertex}), $H_{eff}$ is the Hamiltonian corresponding to $\mathcal{L}_{eff}$ in (\ref{Leff}). We have made the directions $n^{\mu}$ to be slightly space like so that the time ordering and anti-time ordering operators do not affect the Wilson lines.

We next factorize $\Gamma(x)$ according to the manner:
\begin{eqnarray}
\Gamma(x)&\equiv&\sum_{\Gamma_{c},\Gamma_{s}}^{color\quad indices}
\Gamma_{c}(\widehat{\psi}_{n,x}^{(0)},\ldots,
\widetilde{\partial}_{m}^{\phantom{m}m\perp}-ig\widehat{A}_{m,x }^{(0)m\perp})
\Gamma_{s}(Y_{n},Y_{m})(x)
\end{eqnarray}
where
$\Gamma_{c}^{\mu}$ is multi-linear with $\widehat{\psi}_{n,x}^{(0)}$ and $\widehat{A}_{m,x }^{(0)m\perp}$ as physical collinear fields in $\Gamma^{\mu}$ connect to different jets at leading order, Wilson lines that appear in $\Gamma_{s}$ depend on type of partons that appear in $\Gamma_{c}$. We have dropped possible color indices of $\Gamma_{c}$ and $\Gamma_{s}$ for simplicity.

The detected hadrons can locate in the back-to-back region, thus Wilson lines of soft gluons do not necessary to cancel out. In this case, there are only two final jets in the hard sub-diagrams at leading order.(\cite{Collions:1980}) There are only Wilson lines along the directions of initial hadrons or the two final jets in this case. We write the part of $|M|^{2}$ in (\ref{trprob}) that depend on soft gluons as:
\begin{eqnarray}
&&\big<0|\Gamma_{s}^{\dag}(\vec{x})e^{-iH_{s}x^{0}}\Gamma_{s}(\vec{0})|0\big>
\nonumber\\
&\simeq& \big<0|\Gamma_{s}^{\dag}(\vec{x}_{\perp})\Gamma_{s}(\vec{0})|0\big>
\nonumber\\
&\simeq& \big<0|1|0\big>
+\big<0|\Gamma_{s}^{\dag}(\vec{x}_{\perp})\Gamma_{s}(\vec{0})|0\big>|_{\text{two back-to-back final jets}}
\end{eqnarray}
where $H_{s}$ is the Hamiltonian corresponding to $\mathcal{L}_{s}$ in (\ref{Ls}). We can then write $|M|^{2}$ in (\ref{trprob}) as:
\begin{eqnarray}
|M|^{2}
&\simeq &
\lim_{T\to\infty}\sum_{X}\int\frac{\ud^{4}k}{(2\pi)^{4}}\int\ud^{4}xe^{-ik\cdot x}
\nonumber\\
&&
_{-T}\big<p_{1}p_{2}|
e^{i\sum_{n^{\mu}}H_{n}^{(0)}(x^{0}+T)}
\Gamma_{c}^{\dag}(\vec{x})
e^{i\sum_{n^{\mu}}H_{n}^{(0)}(T-x^{0})}
|H_{3}H_{4}X\big>_{T}
\nonumber\\
&&
_{T}\big<H_{3}H_{4}X|
e^{-i\sum_{n^{\mu}}H_{n}^{(0)}T}
\Gamma_{c}(\vec{0})
e^{-i\sum_{n^{\mu}}H_{n}^{(0)}T}
|p_{1}p_{2}\big>_{-T}
\nonumber\\
&&
\left(1
+\big<0|\Gamma_{s}^{\dag}(\vec{x}_{\perp})\Gamma_{s}(\vec{0})|0\big>|_{\text{two back-to-back final jets}}\right)
\end{eqnarray}
where $H_{n}^{(0)}$ is the Hamiltonian corresponding to $\mathcal{L}_{n}^{(0)}$ in (\ref{Ln}).

We then define the annihilation operator:
\begin{eqnarray}
\label{fermion}
\widehat{a}_{n,p}^{s}&=&\frac{\sqrt{2E_{p}}}{2m}\int\ud^{3}\vec{x}_{n}
                             \bar{u}^{s}(p)\widehat{\psi}_{n,x}^{(0)}(\vec{x}_{n})
                             e^{-i\vec{p}\cdot \vec{x}}
\end{eqnarray}
\begin{eqnarray}
\label{antif}
\widehat{a}_{n,p}^{s}&=&-\frac{\sqrt{2E_{p}}}{2m}\int\ud^{3}\vec{x}_{n}
                   \bar{\widehat{\psi}}_{n,x}^{(0)}(\vec{x}_{n})
                   v^{s}(p)e^{-i\vec{p}\cdot \vec{x}}
\end{eqnarray}
for collinear fermions and anti-fermions respectively. One can verify that $\widehat{a}_{n,p}^{s}|0\big>=0$. For collinear gluons,  we define that:
\begin{eqnarray}
\label{gaugeb}
\widehat{a}_{n,p}^{j}&=&\frac{i}{\bar{n}\cdot p}\sqrt{2E_{p}}
\int\ud^{3}\vec{x}_{n}e^{-i\vec{p}\cdot \vec{x}}
\bar{n}\cdot\partial\epsilon^{j*}(p)\cdot\widehat{A}_{n,x}^{(0)}(\vec{x}_{n})
\nonumber\\
&=&\frac{i}{\bar{n}\cdot p}\sqrt{2E_{p}}
\int\ud^{3}\vec{x}_{n}e^{-i\vec{p}\cdot \vec{x}}
\epsilon_{\mu}^{j*}(p)
\nonumber\\
&&
W_{n,x}^{(0)\dag}G_{n}^{(0)n\mu}
W_{n,x}^{(0)}(\vec{x}_{n})
\end{eqnarray}
with $j$ denotes different polarizations, where
\begin{equation}
G_{n}^{(0)n\mu}=\frac{1}{-ig}
[\bar{n}\cdot(\partial-igA_{n}^{(0)}),
\partial^{\mu}-igA_{n}^{(0)\mu}]
\end{equation}
One can also verify that $\widehat{a}_{n,p}^{i}|0\big>=0$. Hadron states can then be expanded according to the parton states $|\widehat{p}_{n}\big>$ which are created by the creation operator $\widehat{a}_{n,p}$. That is:
\begin{equation}
|\widehat{p}_{n}\big>=\sqrt{2E_{p_{n}}}\widehat{a}_{n,p_{n}}^{\dag}|0\big>
\end{equation}
where $\widehat{a}_{n,p_{n}}^{\dag}$ denote the conjugation of the operators (\ref{fermion}), (\ref{antif}) or (\ref{gaugeb}).

$\Gamma$ is multi-linear with $\widehat{\psi}_{n,x}^{(0)}$ and $\widehat{A}_{m,x}^{(0)m\perp\mu}$, where $n\neq m$. There can be no more than one parton states that are created by the operator $\widehat{a}_{n_{i},p}^{\dag}$ contracted with $\Gamma^{\mu}$ for each collinear jets.    We denote these active partons as $|\widehat{p}_{i}^{1}\big>$ and $|\widehat{k}_{i}^{1}\big>$  and have:
\begin{eqnarray}
\label{factorization}
|M|^{2}
&\simeq &
\frac{1}{D(G)}\lim_{T\to\infty}\sum_{X}\int\frac{\ud^{4}k}{(2\pi)^{4}}\int\ud^{4}xe^{-ik\cdot x}
\nonumber\\
&&
\prod_{i=1}^{2}\left(2E_{p_{i}^{1}}
tr_{c}\left\{
\quad
_{-T}\big<p_{i}|e^{iH^{(0)}_{n_{i}}(x^{0}+T)}
\widehat{a}_{n_{i},x,p_{i}^{1}}^{\dag}\right.\right.
\nonumber\\
&&
\left.\left.
e^{-iH^{(0)}_{n_{i}}x^{0}}
\widehat{a}_{n_{i},0,p_{i}^{1}}e^{-iH^{(0)}_{n_{i}}T}
|p_{i}\big>_{T}\right\}\right)
\nonumber\\
&&
\prod_{i=3}^{4}(2E_{k_{i}^{1}}
tr_{c}\left\{
\big<0|
\widehat{a}_{n_{i},x,k_{i}^{1}}
e^{iH^{(0)}_{n_{i}}(T-x^{0})}
|H_{i}X_{n_{i}}\big>_{T}\right.
\nonumber\\
&&
\left.
 _{T}\big<H_{i}X_{n_{i}}|
e^{-iH^{(0)}_{n_{i}}T}
\widehat{a}_{n_{i},0,k_{i}^{1}}^{\dag}
|0\big>\right\}
\nonumber\\
&&
\big<\widehat{p}_{1}^{1}\widehat{p}_{2}^{1}|
\Gamma_{c}^{\dag}(\vec{x})
e^{i\sum_{\widetilde{n}}
H^{(0)}_{\widetilde{n}}(T- x^{0})}
|\ldots \widehat{k}_{i}^{1}\ldots X_{\widetilde{n}}\big>_{T}
\nonumber\\
&&
_{T}\big<\ldots \widehat{k}_{i}^{1}\ldots X_{\widetilde{n}}|
e^{-i\sum_{\widetilde{n}}
H^{(0)}_{\widetilde{n}}T}
\Gamma(\vec{0})|\widehat{p}_{1}^{1}\widehat{p}_{2}^{1}\big>
\nonumber\\
&&
\left(1
+\big<0|\Gamma_{s}^{\dag}(\vec{x}_{\perp})\Gamma_{s}(\vec{0})|0\big>|_{\text{two back-to-back final jets}}\right)
\end{eqnarray}
where $\frac{1}{D(G)}$ denotes the color factor, $X_{n_{i}}$ denotes undetected hadrons that collinear to $n_{i}^{\mu}$,  $\widetilde{n}$ denotes the directions of collinear hadrons that are quite different from those of initial hadrons and detected final hadrons.

We notice that the states $|\widehat{p}_{n}\big>$ is invariant under the gauge transformation $U_{n}(x_{n})$($U_{n}(\infty)=1$):
\begin{equation}
\psi_{n}^{(0)}\to U_{n}\psi_{n}^{(0)},\quad
A_{n}^{(0)\mu}\to U_{n}(A_{n}^{(0)\mu}+\frac{i}{g}\partial^{\mu})U_{n}^{\dag}
\end{equation}
\begin{equation}
\psi_{s}\to \psi_{s},\quad
A_{s}^{\mu}\to A_{s}^{\mu}
\end{equation}
$\Gamma(x)$ and $\Gamma_{c}(x)$ is also invariant under such gauge transformation even if we choose different $U_{n}$ for different directions $n^{\mu}$. Especially, we can choose that $U_{\widetilde{n}}=1$. The matrix-elements between parton states in (\ref{factorization}) are invariant under such gauge transformations.  We can  take the lowest perturbation of the fields $\bar{n_{i}}\cdot A_{n_{i}}^{(0)}$($i=1,\ldots , 4$
) in such matrix-element  as they are scalar polarized.  We then have:
\begin{eqnarray}
|M|^{2}
&\simeq &
\frac{1}{D(G)}\lim_{T\to\infty}\sum_{X}\int\frac{\ud^{4}k}{(2\pi)^{4}}\int\ud^{4}xe^{-ik\cdot x}
\nonumber\\
&&
\prod_{i=1}^{2}\left(2E_{p_{i}^{1}}
tr_{c}\left\{
\quad
_{-T}\big<p_{i}|e^{iH^{(0)}_{n_{i}}(x^{0}+T)}
\widehat{a}_{n_{i},x,p_{i}^{1}}^{\dag}\right.\right.
\nonumber\\
&&
\left.\left.
e^{-iH^{(0)}_{n_{i}}x^{0}}
\widehat{a}_{n_{i},0,p_{i}^{1}}e^{-iH^{(0)}_{n_{i}}T}
|p_{i}\big>_{T}\right\}\right)
\nonumber\\
&&
\prod_{i=3}^{4}(2E_{k_{i}^{1}}
tr_{c}\left\{
\big<0|
\widehat{a}_{n_{i},x,k_{i}^{1}}
e^{iH^{(0)}_{n_{i}}(T-x^{0})}
|H_{i}X_{n_{i}}\big>_{T}\right.
\nonumber\\
&&
\left.
 _{T}\big<H_{i}X_{n_{i}}|
e^{-iH^{(0)}_{n_{i}}T}
\widehat{a}_{n_{i},0,k_{i}^{1}}^{\dag}
|0\big>\right\}
\nonumber\\
&&
\big<p_{1}^{1}p_{2}^{1}|
\Gamma_{c}^{\dag}(\vec{x})
e^{i\sum_{\widetilde{n}}
H^{(0)}_{\widetilde{n}}(T- x^{0})}
|\ldots k_{i}^{1}\ldots X_{\widetilde{n}}\big>_{T}
\nonumber\\
&&
_{T}\big<\ldots k_{i}^{1}\ldots X_{\widetilde{n}}|
e^{-i\sum_{\widetilde{n}}
H^{(0)}_{\widetilde{n}}T}
\Gamma(\vec{0})|p_{1}^{1}p_{2}^{1}\big>|_{\bar{n_{i}}\cdot A_{n_{i}}^{(0)}
=0}
\nonumber\\
&&
\left(1
+\big<0|\Gamma_{s}^{\dag}(\vec{x}_{\perp})\Gamma_{s}(\vec{0})|0\big>|_{\text{two back-to-back final jets}}\right)
\end{eqnarray}
where $|p_{i}^{1}\big>$ is the usual partons produced by the operator $a_{n_{i},p_{i}^{1}}^{\dag}=\widehat{a}_{n_{i},p_{i}^{1}}^{\dag}|_{\bar{n}\cdot A_{n_{i}}=0}$. The condition $\bar{n}\cdot A_{n}=0$ should be treated as the lowest perturbation of the fields $\bar{n}\cdot A_{n}$ not the axial gauge.

If we do not consider the back-to-bak region, then we can set that:
\begin{equation}
x\to \widehat{x}_{n}\equiv(n\cdot x,0,\vec{0})
\end{equation}
in the fields collinear to $n^{\mu}$. In the back-to-back region, we can set that:
\begin{equation}
x\to \widetilde{x}_{n}\equiv(n\cdot x,0,\vec{x}_{n\perp})
\end{equation}
in the fields collinear to $n^{\mu}$. We can then write the transition probability as:
\begin{eqnarray}
\label{final}
|M|^{2}&=&|M|_{c}^{2}+|M|_{b}^{2}
\end{eqnarray}
\begin{eqnarray}
|M|_{c}^{2}
&\equiv&
\frac{1}{D_{c}(G)}\lim_{T\to\infty}\sum_{X}\int\frac{\ud^{4}k}{(2\pi)^{4}}\int\ud^{4}xe^{-ik\cdot x}
\nonumber\\
&&
\prod_{i=1}^{2}\left(2E_{p_{i}^{1}}
tr_{c}\left\{
\quad
_{-T}\big<p_{i}|e^{iH^{(0)}_{n_{i}}(\widehat{x}_{n_{i}}^{0}+T)}
\widehat{a}_{n_{i},\widehat{x}_{n_{i}},p_{i}^{1}}^{\dag}\right.\right.
\nonumber\\
&&
\left.\left.
e^{-iH^{(0)}_{n_{i}}\widehat{x}_{n_{i}}^{0}}
\widehat{a}_{n_{i},0,p_{i}^{1}}e^{-iH^{(0)}_{n_{i}}T}
|p_{i}\big>_{T}\right\}\right)
\nonumber\\
&&
\prod_{i=3}^{4}(2E_{k_{i}^{1}}
tr_{c}\left\{
\big<0|
\widehat{a}_{n_{i},\widehat{x}_{n_{i}},k_{i}^{1}}
e^{iH^{(0)}_{n_{i}}(T-\widehat{x}_{n_{i}}^{0})}
|H_{i}X_{n_{i}}\big>_{T}\right.
\nonumber\\
&&
\left.
 _{T}\big<H_{i}X_{n_{i}}|
e^{-iH^{(0)}_{n_{i}}T}
\widehat{a}_{n_{i},0,k_{i}^{1}}^{\dag}
|0\big>\right\}
\nonumber\\
&&
\big<p_{1}^{1}p_{2}^{1}|
\Gamma_{c}^{\dag}(\vec{x})
e^{i\sum_{\widetilde{n}}
H^{(0)}_{\widetilde{n}}(T- x^{0})}
|\ldots k_{i}^{1}\ldots X_{\widetilde{n}}\big>_{T}
\nonumber\\
&&
_{T}\big<\ldots k_{i}^{1}\ldots X_{\widetilde{n}}|
e^{-i\sum_{\widetilde{n}}
H^{(0)}_{\widetilde{n}}T}
\Gamma(\vec{0})|p_{1}^{1}p_{2}^{1}\big>|_{\bar{n_{i}}\cdot A_{n_{i}}^{(0)}=0}
\end{eqnarray}
\begin{eqnarray}
|M|_{b}^{2}
&\equiv&
\frac{1}{D_{b}(G)}\lim_{T\to\infty}\sum_{X}\int\frac{\ud^{4}k}{(2\pi)^{4}}\int\ud^{4}xe^{-ik\cdot x}
\nonumber\\
&&
\prod_{i=1}^{2}\left(2E_{p_{i}^{1}}
tr_{c}\left\{
\quad
_{-T}\big<p_{i}|e^{iH^{(0)}_{n_{i}}(\widetilde{x}_{n_{i}}^{0}+T)}
\widehat{a}_{n_{i},\widetilde{x}_{n_{i}},p_{i}^{1}}^{\dag}\right.\right.
\nonumber\\
&&
\left.\left.
e^{-iH^{(0)}_{n_{i}}\widetilde{x}_{n_{i}}^{0}}
\widehat{a}_{n_{i},0,p_{i}^{1}}e^{-iH^{(0)}_{n_{i}}T}
|p_{i}\big>_{T}\right\}\right)
\nonumber\\
&&
\prod_{i=3}^{4}(2E_{k_{i}^{1}}
tr_{c}\left\{
\big<0|
\widehat{a}_{n_{i},\widetilde{x}_{n_{i}},k_{i}^{1}}
e^{iH^{(0)}_{n_{i}}(T-\widetilde{x}_{n_{i}}^{0})}
|H_{i}X_{n_{i}}\big>_{T}\right.
\nonumber\\
&&
\left.
 _{T}\big<H_{i}X_{n_{i}}|
e^{-iH^{(0)}_{n_{i}}T}
\widehat{a}_{n_{i},0,k_{i}^{1}}^{\dag}
|0\big>\right\}
\nonumber\\
&&
\big<p_{1}^{1}p_{2}^{1}|
\Gamma_{c}^{\dag}(\vec{x})
e^{i\sum_{\widetilde{n}}
H^{(0)}_{\widetilde{n}}(T- x^{0})}
|\ldots k_{i}^{1}\ldots\big>_{T}
\nonumber\\
&&
_{T}\big<\ldots k_{i}^{1}\ldots|
e^{-i\sum_{\widetilde{n}}
H^{(0)}_{\widetilde{n}}T}
\Gamma(\vec{0})|p_{1}^{1}p_{2}^{1}\big>|_{\bar{n_{i}}\cdot A_{n_{i}}^{(0)}=0}
\nonumber\\
&&
tr_{c}\{
\big<0|\Gamma_{s}^{\dag}(\vec{x}_{\perp})\Gamma_{s}(\vec{0})|0\big>|_{\text{two back-to-back final jets}}\}
\end{eqnarray}
where $|M|_{c}^{2}$ and $|M|_{b}^{2}$ represent contributions of large transverse momenta region and back-to-back region respectively, $\frac{1}{D_{c}(G)}$ and $\frac{1}{D_{b}(G)}$ represent the color factors.

The formula (\ref{final}) is our final result. Soft gluons and scalar polarized gluons that collinear to initial and detected final hadrons decouple from the matrix-element between parton states in (\ref{final}). Such matrix-element can be calculated perturbatively.
\\

\section{Conclusion}
\label{conclusion}

We have presented the proof of TMD-factorization in hadron-hadron collisions in this paper. We constrain that the detected hadrons are not collinear to initial hadrons. This is essential in the cancellation of pinch singular surfaces in Glauber region. Our result contradicts the widely accepted view point that TMD-factorization does not hold for such process. We do not claim that calculations in the literatures are incorrect. We simply point out that there are some missing aspects in these calculations.

The Ward identity cancellation of scalar polarized gluons is prevented by singular points of the type $\bar{n}\cdot l_{n}^{i}=0$ or $\bar{n}\cdot (l_{n}^{1}+\ldots l_{n}^{i})=0$, where $l_{n}^{i}$ represent momenta of scalar polarized gluons that collinear to $n^{\mu}$.  Such singular points indeed locate in soft region. The collinear approximation does not hold in this region. One may first deform the integral path of $\bar{n}\cdot l_{n}^{i}$ to collinear region so that the collinear approximation works.  However, $\bar{n}\cdot l_{n}^{i}$ is pinched in the soft(non-Glauber) region. Thus one can not simply deform the integral path before contributions of the soft region is subtracted from the whole integral.

We first make some subtraction(like that we do in Sec.\ref{cancellation}) so that contributions of the soft region do not affect the remanding part.  This can be written as:
\begin{equation}
M=M_{s}+M-M_{s}\simeq M_{s}+M_{c}
\end{equation}
where $M$ represents the whole amplitude or the modular square of the whole amplitude, $M_{s}$ is achieved by apply some approximation to $M$ which behaves the same as $M$ at leading order in soft region, $M_{c}$ represents that we take the collinear approximation in the $M-M_{s}$ part. The Ward identity cancelation works in $M_{c}$ part as singular points of the type $\bar{n}\cdot l_{n}^{i}=0$ or $\bar{n}\cdot (l_{n}^{1}+\ldots l_{n}^{i})=0$ do not contribute to $M_{c}$ at leading order.

If the gluons exchanged between collinear particles in $M_{s}$ can be absorbed into  Wilson lines along the directions of the collinear particles, then both $M_{c}$ and $M_{s}$ take the factorized form like that shown in Fig.\ref{soft} and Fig.\ref{collinear diagram}. To achieve this, one should first prove the cancelation of pinch singular surfaces in Glauber region so that the Grammer-Yennie approximation works in soft region. This is performed in sec.\ref{deformation}. The deformations depend on wether the collinear particles collinear to initial hadrons or other directions. However, this only affect the Wilson lines of soft gluons.

We see that different deformations of the  integral path can only affect the $M_{s}$ part as $M_{c}$ is free of singular points of the type $\bar{n}\cdot l_{n}^{i}=0$ or $\bar{n}\cdot (l_{n}^{1}+\ldots l_{n}^{i})=0$. For $M_{s}$ part we do need different Wilson lines while gluons couple to different jets. These Wilson lines only depend on which jets the gluons couple to . They do not depend the other ends of the gluon propagators. Thus these Wilson lines can be written into a soft factor after we sum over all undetected final states.

The soft factors do not cancel out if there detected hadrons in back-to-back region. There are only four Wilson lines along the directions of initial hadrons or the two detected hadrons in this case. Thus the soft factor only depend on properties of the initial and detected hadrons. We see that the soft factors do not affect the factorization.

\section*{Acknowledgments}
The author thanks Y. Q. Chen for for helpful discussions and important suggestions about the article.
This work was supported by the National Nature Science Foundation of China under grants No.11275242.

\bibliography{cancellation}

\end{document}